\documentclass[preprint,12pt]{elsarticle}

\usepackage{amssymb}
 \usepackage{amsthm}
 \usepackage{mathrsfs}
  \usepackage{bm}
   \usepackage{caption}
     \usepackage{booktabs}
    \usepackage[namelimits]{amsmath}

\journal{ ArXiv}

\begin{document}

\begin{frontmatter}

\title{Constrained velocity-free control of spacecraft attitude via explicit reference governor}

\author[label1]{Qingqing Dang}

    \affiliation[label1]
    {
    	organization={School of Civil Aviation, Northwestern Polytechnical University},
    	city={Xi'an},
    	postcode={710072}, 
    	state={Shanxi},
    	country={China}
    }   

\tnotetext[label2]{}
\author{Wenbo Li\corref{cor1}\fnref{label2}}
\ead{liwenbo_502@163.com}
\cortext[cor1]{Corresponding author} 
\affiliation[label2]{organization={Beijing Institute of Control Engineering, China Academy of Space  Technology},
	city={Beijing},
	postcode={100190}, 
	country={China}}

  \author[label3]{Haichao Gui}
  \affiliation[label3]{organization={School of Astronautics, Beihang University},
  	city={Beijing},
  	postcode={100191}, 
  	country={China}}

\begin{abstract}
This paper introduces an explicit reference governor-based control scheme tailored for addressing the velocity-free spacecraft attitude maneuver problem. This problem is subject to specific constraints, namely the pointing constraint, angular velocity constraint, and input saturation.
The proposed control scheme operates in two layers, ensuring the asymptotic stability of the spacecraft's attitude while adhering to the aforementioned constraints. 
The inner layer employs output feedback control utilizing an angular velocity observer based on immersion and invariance technology. This observer facilitates attitude stabilization without the measurement of angular velocity.
Through an analysis of the geometry associated with the pointing constraint, determination of the upper bound of angular velocity, and optimization of the control input solution, the reference layer establishes a safety boundary described by the invariant set. 
Additionally, we introduce the dynamic factor related to the angular velocity estimation error into the invariant set to prevent states from exceeding the constraint set due to unmeasurable angular velocity information. The shortest guidance path is then designed in the reference layer.  Finally, we substantiate the efficacy of the proposed constrained attitude control algorithm through numerical simulations.

\end{abstract}


\end{frontmatter}


\section{Introduction}\label{sec1}
Spacecraft attitude maneuver plays a significant role in complex space autonomous missions\cite{ 8424433}. Restricted by the actuators and sensitive payloads, attitude maneuver algorithms must ensure system stability while simultaneously adhering to multiple constraints  \cite{QI2023292, NAKANO2023111103, QU202283}.
For instance, the spacecraft is usually required to maneuver from one state to another within the defined time , while keeping its star sensor avoid from the bright objects (e.g. earth) and preventing the command torque from exceeding the capacity of the actuator \cite{G4469}.
These missions exemplify spacecraft maneuvering amidst state and control constraints \cite{CS, DANG2021}.
Moreover, in scenarios where gyroscopes fail, the unavailability of angular velocity information becomes a significant challenge \cite{POURTAKDOUST2022134}.  Consequently, constrained velocity-free attitude control is an issue of great theoretical and practical importance.

For attitude control systems with actuator  saturation, if the  input constraints are not considered in the    controller design explicitly, although the performance is affected by the input limitation, its stability  sometimes can still be proved theoretically \cite{RN41}.  Traditional controllers designed directly through Lyapunov function  lack the ability to restrict state trajectories.  Hence, the attitude commands  are  used in the attitude maneuver path design in the presence of  multiple  attitude constraints \cite{G4469,   G60189}. This strategy can effectively solve part of the engineering problems, but it suffers from limited flexibility and struggles to meet tasks demanding high real-time dynamic requirements.
Addressing challenges caused by actuator saturation, control bandwidth limits, slew rate constraints, and/or eigenaxis slew constraints, Wie et al. introduced saturation and integration functions within a nonlinear feedback control logic for rapid re-targeting control of agile spacecraft \cite{2002Rapid}. This method can handle single-axis maneuvers with particular constraints well. However, it encounters difficulties in handling three-axis maneuvers with intricate constraints.

The amalgamation of potential functions and Lyapunov functions presents a promising technology for addressing complex constraints in the constrained attitude control problem  \cite{  CHENG201861, kulumani2017constrained, HUA2024108738}.
Lee et al.   \cite{6978863} constructed a strictly convex logarithmic barrier potential for attitude-constrained zones by utilizing a convex parameterization technology. Inspired  by\cite{6978863} and using the anti-unwinding attitude error function,      a new algorithm for the attitude reorientation guidance under forbidden pointing constraints is proposed in  \cite{G003606}.
Furthermore,  Shen et al.  \cite{SHEN2018157} addressed the rest-to-rest three-axis attitude reorientation  under multiple attitude-constraint zones and angular velocity limits via   a quadratic potential function and a logarithmic potential function.
Nevertheless,  it is  difficult to simultaneously handle different types of complex constraints by the potential functions based  constrained control algorithm.
Since the potential function is constructed in the Lyapunov function and the convergence of  Lyapunov function is the result of the convergence game between potential function and states, the robustness of the system may become worse.

Trajectory optimization methods, notably   model predictive control (MPC), offer a means to tackle constrained control issues.   In \cite{KALABIC2017293} and\cite{GJCD_DAE}, MPC on SO(3) has been developed for constrained attitude maneuver of fully actuated spacecraft. However, the necessity to optimize the function at each sampling horizon in MPC restricts its application in systems requiring rapid response, such as spacecraft maneuvering.
Recently, a novel add-on control  scheme  called explicit reference governor (ERG) was introduced by  Nicotra et al. \cite{8412335, 7244340, 8890837, 8675479}. The key idea   is to  augment a pre-stabilized system with a  control unit and  manipulates the auxiliary reference to ensure constraint satisfaction, which means   the stability and the constraint issues   can be handled separately.
This control technology has found application in solving problems related to Unmanned Aerial Vehicles and spacecraft attitude control with state constraints \cite{ NAKANO2023111103, DANG2021,   Nicotra2016ConstrainedCO, CHI2024108874}.

In addition to constraints, another challenge in attitude control is the velocity-free control problem.
This issue has drawn significant attention from researchers and has been extensively studied \cite{HASAN2023,  Yang2016Immersion, ESPINDOLA2022377}. For instance, immersion and invariance (I\&I) technology have been utilized to develop a globally exponentially convergent observer for the angular velocity in \cite{RN41,  Yang2016Immersion, RN42,  6832520}.
In our earlier research, a six-degree-of-freedom observer was constructed using the I\&I approach 
\cite{ G004302}.
However, the velocity-free  attitude maneuver problem in the presence of constraints was studied in just a few  works. For instance,    a velocity-free attitude reorientation control law with  pointing constraints  is established  in  \cite{G002129}.

Inspired by the  ERG and the I\&I technologies, a constrained  velocity-free  control algorithm for spacecraft  reorientation is presented in this paper. The algorithm takes into account attitude pointing, angular velocity, and control input constraints. The formulation of attitude dynamics and various constraints is expressed using modified Rodrigues parameters (MRPs).
The  MRPs constitute a singular,  nonunique and minimal parametrization set of the  three-dimensional special orthogonal group SO(3). Fortunately, the  singularity can be avoided by using the  nonuniqueness properties through switching the parameters between  MRPs and its  shadow at the  unit sphere \cite{AIAAbook, GUI20155832}.
Subsequently,  the ERG-based control scheme is derived, wherein the output controller, relying on the angular velocity observer, is primarily designed in the inner loop. To the best of the authors' knowledge, the result presented in this paper is the first attempt to address the observer-based  attitude maneuver issue with pointing constraints, angular velocity constraints, and input constraints.
Finally, the performance and robustness of the proposed algorithm is verified by the numerical simulations and Monte Carlo simulations.

\section{Preliminaries }
\subsection{Spacecraft attitude kinematics and dynamics}

The   MRPs vector  is defined in terms of  an Euler  rotation angle $  \phi \in\mathbb{R}  $  about the  principal axis $ \{ \textbf{\textit{n}}|\textbf{\textit{n}}^T\textbf{\textit{n}}=1, \textbf{\textit{n}}\in \mathbb{R}^3 \} $.
Let $ \mathscr{F}_B $ be the body-fixed frame, and $ \mathscr{F}_I $ be the inertial frame.
Then, the   attitude with respect to the inertial frame can be described by MPRs and given by $  \boldsymbol{ \sigma}_{BI} =\textbf{\textit{n}}_{BI}{\rm{tan}}(\phi_{BI}/4)  $.
The attitude  kinematics and dynamics of the rigid-body spacecraft  are given by \cite{G002873}

\begin{subequations}\label{Dyn}
	\begin{equation}\label{Dyn_1}
		\dot{\boldsymbol{ \sigma}}_{BI}= G(\boldsymbol{ \sigma}_{BI})\boldsymbol{  \omega}_{BI}^B
	\end{equation}
	$	G(\boldsymbol{ \sigma}_{BI})=\dfrac{1}{2}\left(\dfrac{1-\boldsymbol{ \sigma}^T_{BI}\boldsymbol{ \sigma}_{BI}}{2}\textbf{\textit{I}}_{3}+\boldsymbol{ \sigma}^{\times}_{BI}+\boldsymbol{ \sigma}_{BI}\boldsymbol{ \sigma}^T_{BI} \right) $
	\begin{equation}\label{Dyn_2}
		\textbf{\textit{J}}\dot{\boldsymbol{  \omega}}_{BI}^B+\boldsymbol{  \omega}_{BI}^B\times\textbf{\textit{J}} \boldsymbol{  \omega}_{BI}^B=\boldsymbol{\tau}_c^{B}+\boldsymbol{\tau}_d^{B}
	\end{equation}
\end{subequations}
where $ \boldsymbol{  \omega}_{BI}^B\in \mathbb{R}^3 $ denotes the angular velocity expressed in the body-fixed frame,  $\textbf{\textit{J}}\in\mathbb{R}^{3\times3} $ is the inertia matrix, 
$ \textbf{\textit{I}}_{3} $   denotes the  identity matrix, and  $ ( \textbf{\textit{x}})^\times $ is the $ 3\times3 $   skew-symmetric  cross-product matrix   associated with vector $\textbf{\textit{x}}\in \mathbb{R}^3$. $ \boldsymbol{\tau}_c^{B} $ and  $\boldsymbol{\tau}_d^{B} $ represent the control torque and the disturbance, respectively.

$ \boldsymbol{ \sigma}_{XY} $ denotes the orientation of \textit{X} frame relative to \textit{Y} frame.
$ \omega^X_{YZ}   $ is the angular velocity of \textit{Y} frame relative to \textit{Z} frame expressed in \textit{X} frame.
Then, the relative attitude   between two frames    is defined as
\begin{equation}
	\boldsymbol{ \sigma}_{\scriptscriptstyle XY}=\dfrac{\boldsymbol{ \sigma}_{\scriptscriptstyle YI}(\boldsymbol{ \sigma}^T_{\scriptscriptstyle XI}\boldsymbol{ \sigma}_{\scriptscriptstyle XI}-1)
		+\boldsymbol{ \sigma}_{\scriptscriptstyle XI}(1-\boldsymbol{ \sigma}^T_{\scriptscriptstyle YI}\boldsymbol{ \sigma}_{\scriptscriptstyle YI})-2\boldsymbol{ \sigma}_{\scriptscriptstyle YI}^\times \boldsymbol{ \sigma}_{\scriptscriptstyle XI}}
	{1+\boldsymbol{ \sigma}^T_{\scriptscriptstyle XI}\boldsymbol{ \sigma}_{\scriptscriptstyle XI}\boldsymbol{ \sigma}^T_{\scriptscriptstyle YI}\boldsymbol{ \sigma}_{\scriptscriptstyle YI}
		+2\boldsymbol{ \sigma}^T_{\scriptscriptstyle YI}\boldsymbol{ \sigma}_{\scriptscriptstyle XI}} \nonumber
\end{equation}
and the    dynamics of $  \boldsymbol{ \sigma}_{\scriptscriptstyle XY} $ is given by 

\begin{subequations}
	\begin{equation}
		\dot{\boldsymbol{ \sigma}}_{XY}= G(\boldsymbol{ \sigma}_{XY})\boldsymbol{  \omega}_{XY}^X
	\end{equation}
	\begin{equation}
		\textbf{\textit{J}}\dot{\boldsymbol{  \omega}}_{XY}^X+\boldsymbol{  \omega}_{XI}^X\times\textbf{\textit{J}} \boldsymbol{  \omega}_{XI}^X-\textbf{\textit{J}}(\boldsymbol{  \omega}^X_{XY}\times\boldsymbol{  \omega}^X_{YI}-\textbf{\textit{ C}}^X_Y\dot{\boldsymbol{  \omega}}_{YI}^Y)=\boldsymbol{\tau}_c^{B}
	\end{equation}
\end{subequations}
where $ \boldsymbol{  \omega}^X_{XY}=\boldsymbol{  \omega}^X_{XI}-\boldsymbol{\omega}^X_{YI}$ and  $
\boldsymbol{\omega}^X_{YI}= \textbf{\textit{ C}}^X_Y\boldsymbol{\omega}^Y_{YI}$.
The rotation matrix in terms of the MRPs from $ Y $ frame to  $ X $ frame can be expressed as

\begin{equation}
	\textbf{\textit{ C}}^X_Y=\textbf{\textit{I}}_{3}+\dfrac{8(\boldsymbol{ \sigma}^{\times}_{XY})^2 -4(1-\boldsymbol{ \sigma}^T_{XY}\boldsymbol{ \sigma}_{XY})\boldsymbol{ \sigma}^{\times}_{XY}}
	{(1+\boldsymbol{ \sigma}^T_{XY}\boldsymbol{ \sigma}_{XY})^2}
\end{equation}
The following properties will be frequently used in this paper:
\begin{equation}\label{re_1}
	\boldsymbol{ \sigma}^T_{XY}G(\boldsymbol{ \sigma_{XY}}) =\left(\dfrac{1+\boldsymbol{ \sigma}^T_{XY}\boldsymbol{ \sigma_{XY}}}{4}\right)\boldsymbol{ \sigma_{XY}}^T 
\end{equation}
\begin{equation}\label{re_2}
	G(\boldsymbol{ \sigma}_{XY})^T G(\boldsymbol{ \sigma})= \left(\dfrac{1+\boldsymbol{ \sigma}^T\boldsymbol{ \sigma}}{4}\right)^2 \textbf{\textit{I}}_3  
\end{equation}

According to the description of MPRs in \cite{AIAAbook},  MRPs have geometric singularities  when   $  \phi =\pm360^\circ $, and  it  is   not unique because of the  shadow set, i.e., $ \boldsymbol{ \sigma}=\boldsymbol{ \sigma}^s,  \boldsymbol{ \sigma}^s=-\boldsymbol{ \sigma}/\boldsymbol{ \sigma}^T\boldsymbol{ \sigma}$. Recalling the definition of $ \boldsymbol{ \sigma} $, one knows that $   \|\boldsymbol{ \sigma}\|\leq 1   $ for all $  |\phi|\leq180 ^\circ$. Thus, the spacecraft attitude can be globally parameterized with the shortest principal rotation  by switching the $ \boldsymbol{ \sigma} $ and  $ \boldsymbol{ \sigma}^s $ at the unit sphere  $   \|\boldsymbol{ \sigma}\|= 1   $. Consequently, we stipulate that the magnitude of $ \boldsymbol{ \sigma} $ is bounded  by 1, i.e.,  $   \|\boldsymbol{ \sigma}\|\leq1   $, which is suited to describe any reorientation.

\subsection{State and control constraints}

The pointing constraint, the angular velocity constraint  and the input limitation are considered in this paper.
For the pointing constraint, we suppose the instantaneous angle $\vartheta  $  between a body-fixed unit  vector  $ \textbf{\textit{r}}^B_c $ (such as cameras) and a  inertial constant  unit   vector $ \textbf{\textit{r}}^I_t $  (observed target) should be maintained in  a half-cone angel
$\vartheta_m  $, i.e., $\theta\leq\vartheta_m  $, which is equivalent to
\begin{equation}\label{point_cons}
	\mathcal{C}_p=\left\lbrace (\boldsymbol\sigma_{BI},  \boldsymbol{  \omega}_{BI}^B ): \textbf{\textit{r}}^B_c \cdot \textbf{\textit{r}}^B_t\geq {\rm cos} (\vartheta_m ),  \vartheta_m\in(0,\frac{\pi}{2})
	\right\rbrace
\end{equation}
where $ \textbf{\textit{r}}^B_t=\textbf{\textit{ C}}^B_I\textbf{\textit{r}}^I_t $ is the expression of
$ \textbf{\textit{r}}^I_t  $ in  $ \mathscr{F}_B $.

In consideration of the   payload requirements, the angular velocity constraint is always  exists. Then the constraint set is given by
\begin{equation}\label{velocity_cons}
	\mathcal{C}_{\omega}=\left\lbrace (\boldsymbol\sigma_{BI},  \boldsymbol{\omega}_{BI}^B ):
	\| \boldsymbol{  \omega}_{BI}^B\|\leq  \omega_{\rm  max},~ \omega_{\rm  max}>0
	\right\rbrace
\end{equation}
where  $ \omega_{max}\in \mathbb{R}^+ $ is the maximum angular velocity  amplitude.

The angular momentum exchange devices such as reaction wheels and control moment  gyros are usually used as the  spacecraft attitude control actuators. These devices  may be saturated when the command torque is large. For simplicity, the actuator  constraint is formulated as

\begin{equation}\label{con_cons}
	\mathcal{C}_{\tau}=\left\lbrace (\boldsymbol\sigma_{BI},  \boldsymbol{\omega}_{BI}^B ):
	\| \boldsymbol{\tau}_c^{B}\|\leq \tau_{\rm  max} ,~ \tau_{\rm  max}>0
	\right\rbrace
\end{equation}
where  $  \tau_{\rm  max}\in \mathbb{R}^+ $ is the maximum allowable control torque.

Finally, the dynamic safety margin of the system is the intersection of  the aforementioned three subsets:
\begin{equation}\label{cons}
	\mathcal{C}= \mathcal{C}_p \cap \mathcal{C}_{\omega}\cap \mathcal{C}_{\tau}
\end{equation}

\subsection{Problem statements}

\begin{figure}[h]
	\centering
	\includegraphics[width=0.88\textwidth]{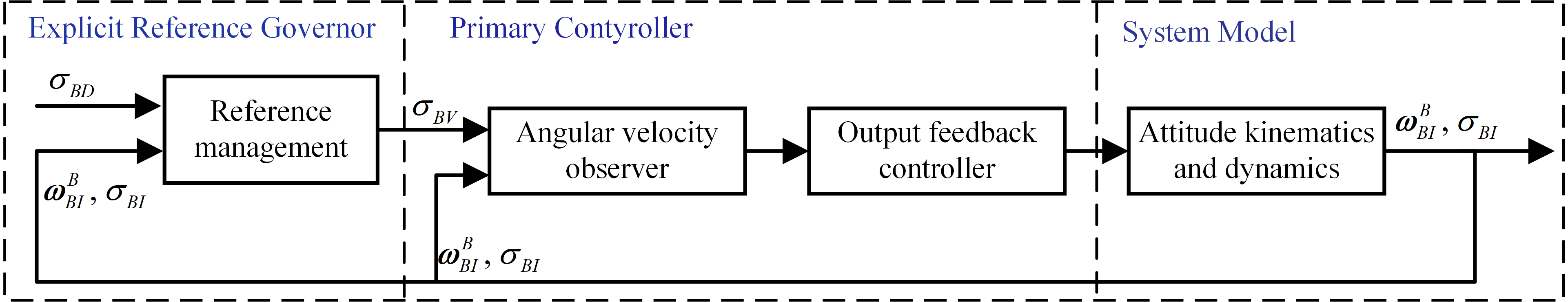}
	\caption{  The   architecture of the explicit reference governor based attitude control  scheme.}\label{ERG}
\end{figure}

This paper aims to develop a ERG control scheme that drives the system states $(\sigma_{BI},  \boldsymbol{  \omega}_{BI}^B)  $   to the desired equilibrium $(\sigma_{DI},  \textbf{0}_{3\times1})  $ while satisfying the constraints \eqref{point_cons}, \eqref{velocity_cons}, and \eqref{con_cons}.
The proposed ERG-based control structure (shown in Fig.  \ref{ERG})  consists of two cascaded control units. The primary controller is given by an angular velocity observer-based output feedback controller, which is able to pre-stabilize the unconstrained system to an auxiliary reference $\sigma_{VI}  $. The reference governor (navigation layer) unit is designed to guarantee the constraint enforcement by manipulating the kinematics of $\sigma_{VI}  $. Clearly, the asymptotic convergence property of the closed-loop control system will be achieved by the   goal that  auxiliary reference $\sigma_{VI}  $  asymptotically tends to $ \sigma_{DI} $.
The reference attitude $\sigma_{DI}$ remains constant and is chosen within the admissible region, i.e., equation \eqref{point_cons} holds true when $\sigma_{BI}= \sigma_{DI}$.


\section{Primary controller design}

This section proposes an angular velocity-free control law so as to stabilize the attitude toward a  constant reference $ \boldsymbol{\overline \sigma}_{VI}   $ ($ \boldsymbol{\dot { \sigma}}_{VI}=\textbf{0}_{3\times1} $) when the constraints and disturbance are neglected ($\boldsymbol{\tau}_d=\textbf{0}_{3\times1} $). The time variation  of $  \sigma_{VI}  $ will be addressed by the reference management unit, which is detailed in the next section.

\subsection{Angular velocity observer design}

The angular velocity observer is constructed based on the  I\&I  theory \cite{  6832520,  G004302}. Let $ \mathscr{F}_E $ be the estimation frame of the  $ \mathscr{F}_B $. The attitude and angular estimation errors in terms of MRPs are given by

\begin{subequations}
	\begin{equation}\label{sigma_error}
		\boldsymbol{ \sigma}_{\scriptscriptstyle BE}=\dfrac{\boldsymbol{ \sigma}_{\scriptscriptstyle EI}(\boldsymbol{ \sigma}^T_{\scriptscriptstyle BI}\boldsymbol{ \sigma}_{\scriptscriptstyle BI}-1)
			+\boldsymbol{ \sigma}_{\scriptscriptstyle BI}(1-\boldsymbol{ \sigma}^T_{\scriptscriptstyle EI}\boldsymbol{ \sigma}_{\scriptscriptstyle EI})-2\boldsymbol{ \sigma}_{\scriptscriptstyle EI}^\times \boldsymbol{ \sigma}_{\scriptscriptstyle BI}}
		{1+\boldsymbol{ \sigma}^T_{\scriptscriptstyle BI}\boldsymbol{ \sigma}_{\scriptscriptstyle BI}\boldsymbol{ \sigma}^T_{\scriptscriptstyle EI}\boldsymbol{ \sigma}_{\scriptscriptstyle EI}
			+2\boldsymbol{ \sigma}^T_{\scriptscriptstyle EI}\boldsymbol{ \sigma}_{\scriptscriptstyle BI}}
	\end{equation}
	\begin{equation}\label{W_error}
		\boldsymbol{  \omega}_{BE}^B=\boldsymbol{  \omega}_{BI}^B-C^B_E\boldsymbol{  \omega}_{EI}^E
		=\boldsymbol{  \omega}_{BI}^B- \boldsymbol{  \omega}_{EI}^B
	\end{equation}
\end{subequations}

To ensure $ \boldsymbol{  \omega}_{BE}^B \rightarrow 0 $ and $ \boldsymbol{ \sigma}_{BE} \rightarrow 0 $, a scalar
$ \varpi\in \mathbb{R}^+$ that can 'cover'  the $ \boldsymbol{  \omega}_{EI}^B $ is introduced as

\begin{equation}\label{W_Cover}
	\varpi=\sqrt{\varepsilon_{\omega}+ \|\boldsymbol{  \omega}_{EI}^B\|^2}
\end{equation}
where $ \varepsilon_{\omega} \in \mathbb{R}^+$ is a constant to be selected, which is utilized to ensure the existence of the time derivative of $  \varpi $.  Then, $ \boldsymbol{  \omega}_{EI}^B $ is generated by
\begin{equation}\label{W_O}
	\boldsymbol{  \omega}_{EI}^B=\boldsymbol{\xi}+4\textbf{\textit{J}}^{-1}\beta(\underline{\varpi } )\boldsymbol{ \sigma}_{BE}
\end{equation}
where $ \underline{\varpi }  $ is the estimate of $  \varpi   $, $ \boldsymbol{\xi} $ and $ \beta(\underline{\varpi } ) $ are the parameter   related to $ \boldsymbol{  \omega}_{EI}^B $ and a function of $ \underline{\varpi } $, respectively. The dynamics of $ \boldsymbol{\xi} $, $  \underline{\varpi }   $, and $ \boldsymbol{ \sigma}_{EI} $
are designed as

\begin{subequations}\label{W_O_detail}
	\begin{align}
		\dot{\boldsymbol{\xi} } =&\textbf{\textit{J}}^{-1}(-\boldsymbol{  \omega}_{EI}^B\times\textbf{\textit{J}}\boldsymbol{  \omega}_{EI}^B+\boldsymbol{\tau}_c^{B})-4\textbf{\textit{J}}^{-1}\dot\beta(\underline{\varpi } )\boldsymbol{ \sigma}_{BE} \notag \\
		&-4\textbf{\textit{J}}^{-1}\beta(\underline{\varpi } )\dot{ \boldsymbol{ \sigma}}_{1BE} \label{W_O1}\\
		\dot {\underline{\varpi }}=\varpi^{-1}& (\boldsymbol{  \omega}_{EI}^B)^T \textbf{\textit{J}}^{-1}
		(-\boldsymbol{  \omega}_{EI}^B\times\textbf{\textit{J}}\boldsymbol{  \omega}_{EI}^B+\boldsymbol{\tau}_c^{B})
		-K_{\underline{\varpi }}(\underline{\varpi }- \varpi  ) \label{W_O2}\\
		&\dot{\boldsymbol{ \sigma}}_{EI}=G(\boldsymbol{ \sigma}_{EI})(\boldsymbol{  \omega}_{EI}^E+K_\sigma \textbf{\textit{C}}^E_B\boldsymbol{ \sigma}_{BE}  )\label{W_O3}
	\end{align}
\end{subequations}
where $  K_{\underline{\varpi }}  $  and $ K_\sigma  $ are the dynamic gains to be designed, $  \dot{ \boldsymbol{ \sigma}}_{1BE}  $ represents part of  the dynamics of $ \boldsymbol{ \sigma}_{BE} $, and it can be obtained from \eqref{Dyn_1},     \eqref{W_error}, and \eqref{W_O3} :

\begin{align}
	\dot{ \boldsymbol{ \sigma}}_{BE} =&G(\boldsymbol{ \sigma}_{BE})(\boldsymbol{\omega}_{BI}^B- \boldsymbol{  \omega}_{EI}^B)\notag\\
	=&G(\boldsymbol{ \sigma}_{BE})[\boldsymbol{\omega}_{BI}^B-\textbf{\textit{C}}^B_E(\boldsymbol{  \omega}_{EI}^E+K_\sigma \textbf{\textit{C}}^E_B\boldsymbol{ \sigma}_{BE}  )]\notag\\
		=&G(\boldsymbol{ \sigma}_{BE})(\boldsymbol{\omega}_{BE}^B-K_\sigma \boldsymbol{ \sigma}_{BE}  )\notag\\
	=&\dot{ \boldsymbol{ \sigma}}_{1BE}+\dot{ \boldsymbol{ \sigma}}_{2BE}\label{segma_err}
\end{align}
with $  \dot{ \boldsymbol{ \sigma}}_{1BE}=-G(\boldsymbol{ \sigma}_{BE})K_\sigma\boldsymbol{ \sigma}_{BE} $ and
$ \dot{ \boldsymbol{ \sigma}}_{2BE}=G(\boldsymbol{ \sigma}_{BE})\boldsymbol{  \omega}_{BE}^B $.

To inject the nonlinear terms in the dynamics of  $ \boldsymbol{  \omega}_{BE}^B $
a dynamic scaling technique is introduced:
\begin{equation}\label{z}
	\textbf{\textit{z}}=\dfrac{\boldsymbol{  \omega}_{BE}^B}{\textit{r}}
\end{equation}
where  $ r $ is the dynamic scaling factor and is updated by the following  law

\begin{equation}\label{r_dyn}
	\dot{\textit{r}}=\dfrac{\textit{r}}{J_m}(J_M\|  \varpi  -\underline{\varpi } \|)-\dfrac{k_r}{J_M}(r-1)
\end{equation}
where $ k_r  \in \mathbb{R}^+$ is the dynamic scaling gain to be determined, $ J_m $ and $ J_M $ are the minimum   and the maximum eigenvalues of the inertia matrix $ \textbf{\textit{J}} $, respectively. 
If  $ r(t)=1 $,  $\dot{\textit{r}}\geq 0$. Hence, it satisfies  $ r(t)\geq1 $ for all $ t\geq0 $ when  $ r(0)\geq1 $. Finally, the convergence analysis of the proposed observer \eqref{W_O} is summarized as the following proposition.

\newtheorem{pro}{proposition}
\begin{pro} \label{proposition1}
	Consider the angular velocity observer in \eqref{W_O} with dynamics given in  \eqref{W_O_detail}, \eqref{r_dyn}, and the gains are given as
	\begin{subequations}\label{obser_gain}
		\begin{equation}\label{obser_gain1}
			\beta(\underline{\varpi } )=4\underline{\beta}(\underline{\varpi } )G^T(\boldsymbol{ \sigma}_{BE})
		\end{equation}
		\begin{equation}\label{obser_gain2}
			\underline{\beta}(\underline{\varpi } )=J_M\| \underline{\varpi }  \|+\dfrac{J_mk_r}{J_M}+1+\rho_{\varpi }
		\end{equation}
		\begin{equation}
			K_{\underline{\varpi }}=8\left(\dfrac{\|\boldsymbol{  \omega}_{EI}^B\|\beta(\underline{\varpi } )r}{J_m}\right)^2
			+\frac{1}{2}r^2J_M+\rho_{\underline{\varpi } }
		\end{equation}
		\begin{equation}
			K_\sigma=\frac{1}{2}r^2+\rho_\sigma
		\end{equation}
		\begin{equation}
			k_r=\frac{1}{2}\dfrac{J^2_M}{J_m}+\rho_r
		\end{equation}
	\end{subequations}
	where $ \rho_{\varpi }  $, $ \rho_{\underline{\varpi } } $, $\rho_\sigma  $, and $ \rho_r $  are positive constants that can be tuned for different convergence rates of the estimation errors.
	Then,  the dynamic scaling factor $ r $ is bounded and  the errors   globally exponentially converges to the origin, i.e.,  ${\rm lim}_{t\rightarrow \infty}e^{\alpha t}\|\boldsymbol{  \omega}_{BE}^B\| =0, \alpha\in \mathbb{R}^+$.
\end{pro}

\textbf{Proof}: 	See the Appendix.
$\hfill\blacksquare$

\subsection{Velocity-free controller design}

The aforementioned angular velocity observer is used to derive a velocity-free feedback attitude controller.
As shown in Fig.  \ref{ERG}, the following theorem summarize the result on the unconstrained output controller.
\newtheorem{thm}{Theorem}

\begin{thm} \label{thm1}
	Consider the attitude dynamics given in \eqref{Dyn} and the angular velocity observer given in \eqref{W_O}-\eqref{obser_gain}. Then,  the   output feedback control law is given by
	\begin{equation}\label{tc_1}
		\boldsymbol{\tau}_c^{B}=-k_p  \boldsymbol{  \sigma }_{BV}-k_d\boldsymbol{  \omega}_{EI}^B
	\end{equation}
	with $ k_p, ~ k_d >0 $, the equilibrium   $ (\boldsymbol{\overline \sigma}_{VI} , \textbf{0}_{3\times1}) $ is asymptotically stable within the admissible set, i.e.,  ${\rm lim}_{t\rightarrow \infty} (\boldsymbol{ \sigma}_{BI}, \boldsymbol{  \omega}_{BI}^B)=(\boldsymbol{\overline \sigma}_{VI} , \textbf{0}_{3\times1}) $.
\end{thm}

\textbf{Proof}:
By employing  \eqref{W_error}, the control law    \eqref{tc_1} can be expressed as a full-state controller augmented by perturbations resulting from velocity estimation errors, namely,
\begin{equation}\label{tc_2}
	\boldsymbol{\tau}_c^{B}=-k_p  \boldsymbol{  \sigma }_{BV}-k_d \boldsymbol{  \omega}_{BI}^B+k_d\boldsymbol{  \omega}_{BE}^B
\end{equation}
Consider a Lyapunov function candidate as follows:
\begin{equation}\label{vc}
	V_c=2k_p{\rm ln}(1+\boldsymbol{ \sigma}^2_{BV})+\frac{1}{2}(\boldsymbol{  \omega}_{BI}^B)^T\textbf{\textit{J}}\boldsymbol{  \omega}_{BI}^B
\end{equation}
Taking the time derivative of   \eqref{vc} along   \eqref{Dyn},   \eqref{re_1},  and  \eqref{tc_2}, one can obtain

\begin{align}
	\dot V_c=&4k_p\dfrac{ \boldsymbol{ \sigma}_{BV}^T\boldsymbol{\dot \sigma}_{BV}}{1+\boldsymbol{ \sigma}_{BV}^2}
	+ (\boldsymbol{  \omega}_{BI}^B)^T\textbf{\textit{J}}\boldsymbol{  \dot\omega}_{BI}^B \notag \\
	\leq&-k_d\|\boldsymbol{  \omega}_{BI}^B\|^2+k_d\|\boldsymbol{  \omega}_{BI}^B\|\|\boldsymbol{  \omega}_{BE}^B\|\label{d_vc}
\end{align}

Clearly, $ \dot V_c $ contains a term with indefinite sign induced by angular velocity estimation error. To eliminate this effect, consider a positive definite Lyapunov function in the following form:

\begin{equation}
	V= V_c+ \delta_zV_z
\end{equation}
where $ \delta_z $ is a positive constant to be determined. Differentiating  $ V $ and applying  \eqref{d_vc} and \eqref{d_Vz} yields
\begin{equation}\label{dot_V}
	\begin{aligned}
		\dot V\leq&-k_d\|\boldsymbol{  \omega}_{BI}^B\|^2+k_d\|\boldsymbol{  \omega}_{BI}^B\|\|\boldsymbol{  \omega}_{BE}^B\|
		-\delta_z(1+\rho_{\varpi })\| \textbf{\textit{z}} \|^2\\
		\leq& -[		\|\boldsymbol{  \omega}_{BI}^B\| 
		\|\boldsymbol{  \omega}_{BE}^B\|	]\!\!
		\begin{bmatrix}
			k_d \!\!\! &-\dfrac{1}{2}k_d \\
			-\dfrac{1}{2}k_d \!\!\! &\delta_zr^{-2} (1+\rho_{\varpi })
		\end{bmatrix}\!\!
		\begin{bmatrix}
			\|\boldsymbol{  \omega}_{BI}^B\| \\
			\|\boldsymbol{  \omega}_{BE}^B\|
		\end{bmatrix}\!
	\end{aligned}
\end{equation}

Given that $1 \leq r < \infty$, there exists a sufficiently large $ \delta_z $ such that $ \dot V $ is negative semi-definite for $ \boldsymbol{\overline \sigma}{VI}$.
By using the LaSalle invariance principle, one can conclude that the equilibrium point $ (\boldsymbol{\overline \sigma}_{VI} , \textbf{0}_{3\times1}) $ of the system is   asymptotically stable. This completes the proof.
$\hfill\blacksquare$

Obviously, $ \boldsymbol{ \sigma}_{VI} $ is time varying,  \textbf{Theorem 1} addresses the claim concerning the tracking error stability of $ \boldsymbol{\overline \sigma}_{VI}$ rather than $ \boldsymbol{ \sigma}_{VI}$.
In fact, since the final state $ \boldsymbol{  \sigma}_{DI}$ is a constant attitude, the inner loop controller only needs to ensure that the attitude can converge to the final state.
Additionally, the precise acquisition of angular velocity is challenging, resulting in difficulties in strictly guaranteeing the angular velocity constraint. Fortunately, the value of the  dynamic scaling factor $ r $  reflects the estimation error. By designing functions associated with $ r $, it becomes possible to satisfy the angular velocity constraints. These properties will be utilized in the subsequent section.

\newdefinition{rmk}{Remark}
\begin{rmk}
	Throughout  the preceding analysis, it is evident that the design of the output feedback controller remains independent of the angular velocity observer (refer to \eqref{tc_1} and \eqref{tc_2}), which greatly reduces the difficulty of the controller design.
	Moreover,    \eqref{W_O_detail} and \eqref{obser_gain} suggest a conflict between the scale of the observer gains and system robustness.   However, there are always parameter uncertainties in practical  missions, so it is necessary to balance the two properties, which will be verified in detail in the simulation.
\end{rmk}

\section{Reference management}
The reference management layer of ERG (illustrated in Fig.  \ref{ERG}) devises  an auxiliary control law that manipulates the reference state  towards the primary stabilized system \cite{661611, 8412335}.
The objective of this section is to handle the constraints outlined in   \eqref{point_cons} $- $ \eqref{con_cons}
by designing the safety margin and the navigation field, which is achieved  by the invariant   set in the  Lyapunov function centered on the reference state  $ \sigma_{VI} $. The formulation of the auxiliary reference takes the following form:
\begin{equation}\label{dot_VI}
	\dot\sigma_{VI}=\Delta(\sigma_{BV},\boldsymbol{  \omega}_{EI}^B)\chi(\sigma_{VI},\sigma_{VD})
\end{equation}
where $ \Delta(\sigma_{BV},\boldsymbol{  \omega}_{EI}^B):\mathbb{R}^3\times\mathbb{R}^3\rightarrow\mathbb{R} $
is the dynamic safety margin that indicates how safe it is  within the allowable set.
$ \chi(\sigma_{VI},\sigma_{VD}): \mathbb{R}^3\times\mathbb{R}^3\rightarrow\mathbb{R}^3 $ denotes the navigation field   of the current state $ \sigma_{VI} $, and the $ \sigma_{VD} $  is utilized to drive the  $ \sigma_{VI} $ towards to $ \sigma_{DI} $.

\subsection{Safety margin}

From an intuitive perspective,  the  safety margin can be treated as the distance between the   constraint boundary and the navigation field. Since $\dot V $ is negative semi-define (see \eqref{dot_V}), the forward invariant set $ \left\lbrace   (\boldsymbol\sigma_{BI},  \boldsymbol{  \omega}_{BI}^B ):  V\leq\Gamma    \right\rbrace  $ can be used to design the safety margin, where  the upper bound $ \Gamma(\boldsymbol{  \sigma}_{VI}^V,\boldsymbol{  \omega}_{VI}^V)  $ is determined by the constraints   \eqref{point_cons}  $- $ \eqref{con_cons}.
In    \cite{   7244340} and \cite{ 8890837}, authors
design the dynamic safety margin in the form $ \Delta(\sigma_{BV},\boldsymbol{  \omega}_{EI}^B)=k_e(\Gamma-V) $, where the constant $ k_e $ is  used to adjust the dynamic performance. Unfortunately, since the exact estimation error $   \boldsymbol{  \omega}_{BE}^B $ is unavailable, the angular velocity   can not be obtained either. In order to prevent $ \Delta $ being negative caused by $   \boldsymbol{  \omega}_{BE}^B $,  $ \Delta $ can be designed as
\begin{equation} \label{deta}
	\Delta(\sigma_{BV}, \boldsymbol{  \omega}_{EI}^B)=\left\{ \begin{aligned}
		k_e(\Gamma-V),~ & ~\Gamma>V\\
		0~~~~~~~,~ &~ \Gamma \leq V
	\end{aligned}
	\right.
\end{equation}

\subsubsection{Pointing constraint}

\begin{figure}[!h]
	\centering
	\includegraphics[width=0.65\textwidth]{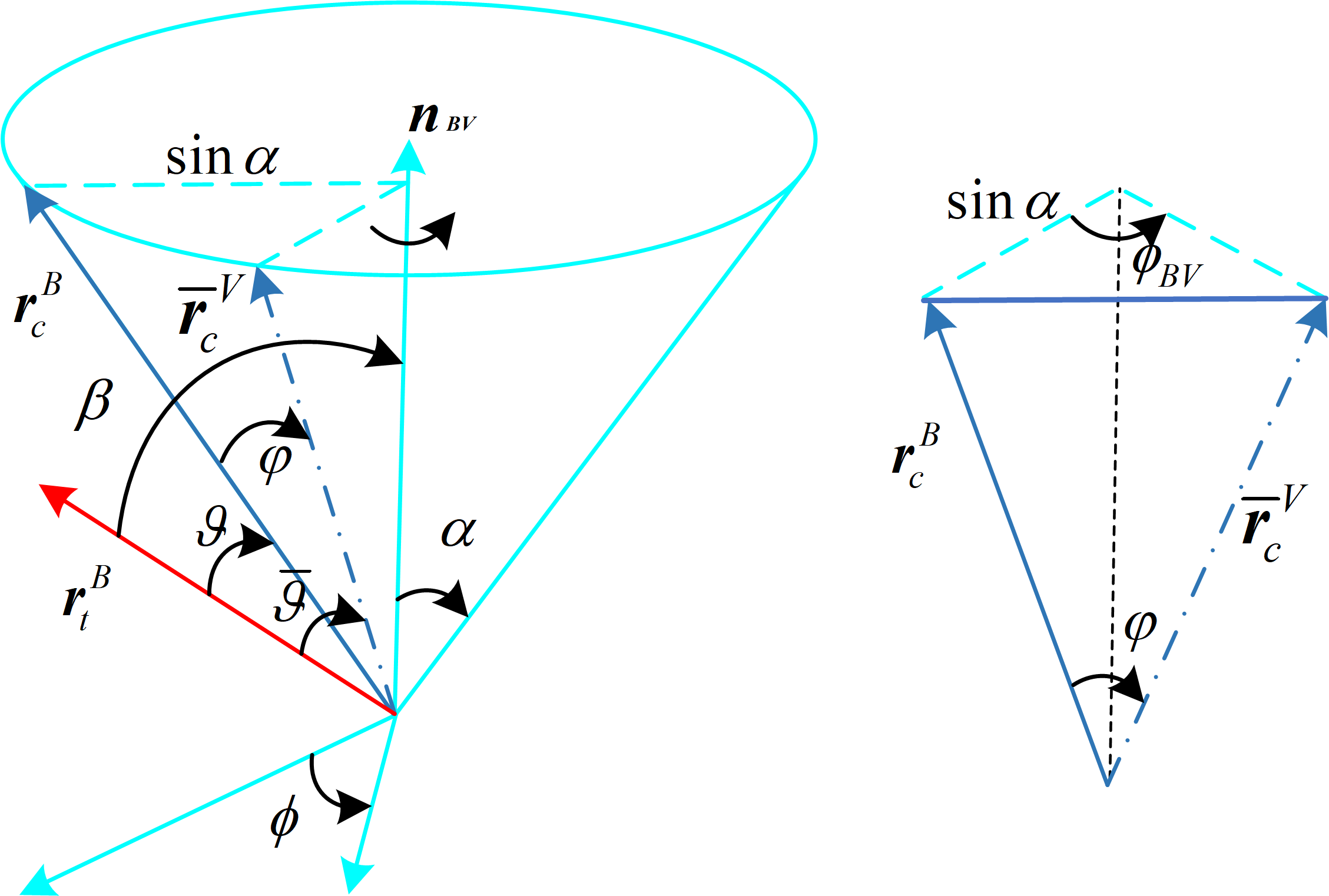}
	\caption{Constrained attitude region.}\label{attitude_semo}
\end{figure}

The geometric relationship about the pointing constraint is displayed in Fig.  \ref{attitude_semo}, where $ \vartheta_e\in(0,\frac{\pi}{2}) $ is the safety margin of   $ \sigma_{BV}^B $ and   satisfies
$  \vartheta_e=\vartheta_m-\vartheta  $.
When the body  frame is coincided with the  reference frame $ \mathscr{F}_V $, the pointing angle $ \overline \vartheta $  is denote by
\begin{equation}
	\overline \vartheta=arcos(\overline{\textbf{\textit{ r}}}^V_c\cdot\textbf{\textit{ r}}^V_t)
\end{equation}
Note that $ \overline{\textbf{\textit{ r}}}^V_c=\textbf{\textit{ r}}^B_c $ is a virtual constant unit vector expressed in $ \mathscr{F}_V $ rather than $ \overline{\textbf{\textit{ r}}}^V_c=\textbf{\textit{C}}^V_B\textbf{\textit{ r}}^B_c $. Under these conditions, the  safety margin  of $ \sigma_{VI}^V $   satisfies
$ \overline\vartheta_e=\vartheta_m-\overline\vartheta  $
and $ \overline\vartheta_e\in(0,\frac{\pi}{2}] $. 
Let $ \alpha\in [0,\pi] $ be the gap between   $ \textit{\textbf{n}}_{BV} $ and the unit vector $  \textbf{\textit{ r}}^B_c  $,
then it satisfies
\begin{equation}
	\textbf{\textit{ r}}^B_c\cdot \textit{\textbf{n}}_{BV}={\rm cos}( \alpha)
\end{equation}

Let $ \varphi\in(0, \pi ) $ denote the orientation  from $ \textbf{\textit{ r}}^B_c $ to $  \overline{\textbf{\textit{ r}}}^V_c $,and they have the following relationship:
\begin{equation}\label{P_erfa}
	{\rm sin}(\dfrac{\varphi}{2})={\rm sin}(\dfrac{\phi_{BV}}{2}  ){\rm sin}(\alpha)
\end{equation}
Obviously, there is a positive correlation between   $ \varphi $ and $\phi_{BV}  $, and they satisfies $ \varphi \leq \phi_{BV}  $.
Since $  | \vartheta_e -\overline\vartheta_e  |\leq  \varphi $, if $ \varphi\leq \overline\vartheta_e  $, then  $ \vartheta_e\geq0 $  can be guaranteed.

According to   \cite{DANG2021}, if $ \boldsymbol{  \omega}_{EI}^E=\textbf{0}_{3\times1} $ ($ \boldsymbol{  \omega}_{BI}^B $ is precisely known), then   $ \dot V_c= \dot V \leq0 $, the threshold of $  V\leq\Gamma_p' $ can be designed as

\begin{equation}
	\Gamma_p'=\left\{ \begin{aligned}
		2k_p{\rm ln} \left\lbrace
		1+\left( \frac{1-\sqrt{1-a_p^2}}{a_p}
		\right)^2
		\right\rbrace &~~\alpha\in(0,\pi)\\
		\infty~~~~~~~~~~~~~~~~~~~~& ~~\alpha=0 ~ {\rm or}  ~\pi
	\end{aligned}
	\right.
\end{equation}
where  $ a_p={\rm sin}(\dfrac{\overline\vartheta_e }{2})/{\rm sin}(\alpha) $.

When $ V=\Gamma_p' $,   $ \dot\sigma_{VI}=\textbf{0}_{3\times1} $. According to  \textbf{Theorem1}, the time derivate of $ V$ and $\Gamma_p' $ satisfies $ \dot V \leq0,  {\dot \Gamma}'_p=0 $, which means the pointing constraint \eqref{point_cons} will never be violated.
However, $ \boldsymbol{  \omega}_{BI}^B $ and $ \delta_z $ are unavailable,  which means $ V $ and  $ V_c $ are unavailable, thus they can not be used to design the threshold of the pointing constraint. Hence, we approximate the Lyapunov function \textit{V} with the following form
\begin{equation}\label{V_like}
	\overline V=  2k_p{\rm ln}(1+\boldsymbol{ \sigma}_{BV})+\frac{1}{2}(\boldsymbol{  \omega}_{EI}^B)^T\textbf{\textit{J}}\boldsymbol{  \omega}_{EI}^B
\end{equation}
Although $\dot {\overline V} $ is sign indefinite, according \textbf{Theorem 1},  $  \overline V $ is asymptotically convergent, i.e.,  ${\rm lim}_{t\rightarrow \infty}   \overline V(t)=0 $. Accordingly, the threshold of the pointing constraint is designed as
\begin{equation}
	\Gamma_p=\dfrac{	\Gamma_p'}{r^{k_1}}
\end{equation}
where $ k_1>0 $ is a constant parameter.  Similarly to the analysis in   Sec. 3.1,
the larger $ \boldsymbol{  \omega}_{BE}^B $ is, the lager $ r $ is, and the smaller $ \Gamma_p $ is.
Although the exact  relationship between $ \boldsymbol{  \omega}_{BE}^B $ and $ r $ is unknown, by tuning $ k_1 $, a conservative but safe threshold of the pointing constraint without angular velocity measurement  can be obtained.

\subsubsection{Angular velocity constraint}

The angular velocity constraint given in  \eqref{velocity_cons}  is a  convex set. As discussed in   Sec. 4.1.1, when $ \boldsymbol{  \omega}_{EI}^E=\textbf{0}_{3\times1} $, the threshold of the angular velocity constraint $  \Gamma_{\omega}  $ can be selected as

\begin{equation}
	\Gamma'_{\omega} =\frac{1}{2}J_m\omega_{\rm max}^2
\end{equation}
Similar  to the pointing constraint, when $ V= \Gamma'_{\omega}$,  $ \dot\sigma_{VI}=\textbf{0}_{3\times1} $. Recalling  \textbf{Theorem1}, the time derivate of them satisfies $ \dot V \leq0,  {\dot \Gamma}'_{\omega}=0 $, which means the angular velocity constraint \eqref{velocity_cons} will never be violated.
When $ \boldsymbol{  \omega}_{EI}^E\neq\textbf{0}_{3\times1} $, the Lyapunov function is approximate by   \eqref{V_like}, and the threshold can be selected as
\begin{equation}
	\Gamma_{\omega} =\frac{	\Gamma'_{\omega}}{  r^{k_2}}
\end{equation}
where $ k_2>0 $ is a constant parameter   used to tuning  $  \Gamma_{\omega} $.

\subsubsection{Actuator saturation}

Similar to the preceding constraints,  $ \boldsymbol{  \omega}_{BI}^B $ is substituted with $ \boldsymbol{  \omega}_{EI}^B $ and we omit the estimation error. Adhering to the approach outlined in  \cite{8890837}, the saturation constraint \eqref{con_cons} can be satisfied by solving the subsequent  optimization problem

$ \textbf{Problem} $
$${\rm min} ~ 2k_p{\rm ln}(1+\boldsymbol{ \sigma}_{BV})+\frac{1}{2}(\boldsymbol{  \omega}_{EI}^B)^T\textbf{\textit{J}}\boldsymbol{  \omega}_{EI}^B $$

subject to
\begin{subequations}
	\begin{equation}
		\lvert\boldsymbol{ \sigma}_{BV} \rvert_i\leq1
	\end{equation}	
	\begin{equation}
		\lvert k_p  \boldsymbol{  \sigma }_{BV}+k_d\boldsymbol{  \omega}_{EI}^B \rvert_i\geq\tau_{\rm  max}
	\end{equation}			
\end{subequations}

Then the threshold $ \Gamma_\tau  $ can be obtained by selecting  the minimum value from the aforementioned optimization problem for $ i={1,2,3} $. Consequently, the  upper-bound of the system   subject to the constraint  \eqref{cons} can be concluded as $ \Gamma={\rm min}\{\Gamma_p,\Gamma_\omega, \Gamma_\tau \} $, a conclusion that can be substantiated by employing the same arguments presented in  \cite{7244340}.

\subsection{Navigation layer}

The navigation field $  \chi(\sigma_{VI},\sigma_{VD}) $ will be designed in this section to ensure that the auxiliary reference $ \boldsymbol{  \sigma }_{BV} $  towards to the desired reference $ \boldsymbol{  \sigma }_{BD} $. Consequently, the trajectory of $  \chi(\sigma_{VD}) $  must strictly align within the permissible set $ 	\mathcal{C} $.
Given that both  the initial  and final attitudes  fall within  the constraints and the pointing constraint $ 	\mathcal{C}_p $  constitutes a convex set, the shortest distance on the attitude manifold adheres to the constraints.
The navigation trajectory $  \chi(\sigma_{VD}) $ is designed by

\begin{equation}\label{nagivation}
	\chi(\sigma_{VD})=-G(\boldsymbol{ \sigma}_{VD}) \boldsymbol{ \sigma}_{VD}
\end{equation}
Since $ \boldsymbol{  \omega}_{BI}^B=\textbf{0}_{3\times1} $  and  $ \boldsymbol{\tau}_c^{B} =\textbf{0}_{3\times1}$ represent the equilibrium point, the constraints \eqref{velocity_cons} and \eqref{con_cons} are always satisfied at steady-state. Then, the main results about the constrained attitude maneuver control without angular velocity measurement is presented in the following proposition.

\begin{pro} Given the spacecraft attitude   dynamics \eqref{Dyn} subject to the constraints \eqref{cons} with the angular velocity observer  \eqref{W_O} controlled by \eqref{tc_1}, and let \eqref{dot_VI} be the navigation layer subject to the dynamic safety margin \eqref{deta}, and the navigation field
	\eqref{nagivation}. Then, for any initial states  satisfy the constraints and $ V(0)\leq \Gamma(0)$, the following statements hold.
	1) For any constant  reference $ \boldsymbol{ \sigma}_{DI}\in	\mathcal{C} $, the system constraints are all satisfied.
	2) The auxiliary reference $ \boldsymbol{ \sigma}_{VI} $  updated by \eqref{dot_VI} asymptotically converges
	to $ \boldsymbol{ \sigma}_{DI} $.
\end{pro}

$  \textbf{Proof}$: See \cite{DANG2021}.
$\hfill\blacksquare$

\begin{rmk}
By integrating the ERG design process with the aforementioned analysis, it can be seen that by designing a trajectory from the current state to the final state that satisfies the constraint conditions, and then the controller drives the system state and the reference state error within a certain range, the system states can be guaranteed to  reach the target state while the constraints are met. An additional advantage of this strategy lies in its capability to maintain effective control even in the absence of state constraints.  In comparison to alternative control algorithms like PID, the ERG algorithm can track the $ {\sigma}_{VD} $    independently generated by the reference management with smaller error than $ {\sigma}_{BD} $. This capability leads to improved control performance characterized by enhanced speed and precision.
\end{rmk}

\section{Numerical simulations}

This section  aims to  demonstrate the effectiveness  of the proposed angular velocity free  attitude control algorithm in the presence of  multi-constraints.
 The objective involves maneuvering the rigid spacecraft from a specific initial state to a pre-defined target, incorporating considerations for attitude constraints, angular velocity constraints, and control saturation concurrently. 
  Besides,  Monte Carlo results are conducted to further verify the  robustness of the proposed control scheme.
The   inertia of the spacecraft is given by
\begin{equation*}\nonumber
	\textbf{\emph{J}}=\begin{bmatrix}15.2&-1&2\\
		-1&18.3&-0.5\\
		2&-0.5&16.1
	\end{bmatrix}\mathrm{kg.m}^{2}.\nonumber
\end{equation*}

The initial states are set as 	$ \boldsymbol{ \sigma}_{BI}(0)= [  -0.119, 0.000, 0.159 ]^T$  and  		$ \boldsymbol{  \omega}(0)=[0, -0.01, 0.01]^T{\rm rad/s} $.
The constraint conditions and target state are chosen in Table \ref{tab1}. Besides, the threshold of actuator saturation $ \Gamma_\tau  $ is obtained by solving from the $ Problem $ via $ fmincon $ function in Matlab 2021, which is 0.0468, and the observer parameters are shown in  Table  \ref{table2}. The control elements are selected as  $  k_p=1.5    $   and $  k_d=2.5    $.
For the  brevity and intuition, Euler angles $ [\varphi, \vartheta,  \psi  ]^T $ with sequence $ 3-1-2 $ are used to plot the  attitude.

\begin{table}[!h]
	\centering
	\captionof{table}{System  Constraint Conditions.\label{tab1}}%
	\begin{tabular*}{240pt}{@{\extracolsep\fill}lcccc@{\extracolsep\fill}}
		\toprule
		\textbf{Parameters} & \textbf{	Values}  \\
		\midrule
		$  \boldsymbol{ \sigma}_{DI} $   &	 $  [ 0, 0, 0 ]^T $ \\
		$ \vartheta_m $     &	  $38^\circ $ 	\\
		$ \textbf{\textit{r}}^I_t $     &	 $[1/\sqrt{3}, -1/\sqrt{3}, 1/\sqrt{3} ]^T $ 	\\
		$ \textbf{\textit{r}}^B_c $  & $ [0, -1/\sqrt{2},   1/\sqrt{2} ]^T $   	   \\
		$ \omega_{max} $           &        $ 0.035  {\rm rad/s}$       \\
		$    \tau_{\rm  max}    $         &    $  0.1$  N.m.s         \\
		$ 	k_e   $     & $ 1000 $  \\
		$ 	k_1, 	k_2    $     & $ 2 $  \\
		\bottomrule
	\end{tabular*}
\end{table}

\begin{table} [!h]
	\centering
	\captionof{table}{Observer parameters.\label{table2}}%
	\begin{tabular*}{240pt}{@{\extracolsep\fill}lcccc@{\extracolsep\fill}}
		\toprule
		\textbf{Parameters} & \textbf{	Values}  \\
		\midrule
		$ J_m	 $	   &$ 18.3 $	   \\
		$ J_M $   &	 $  15.2 $ \\
		$ \rho_\sigma , \rho_{\varpi },  \rho_{\underline{\varpi } },\rho_r , \varepsilon_{\omega}  $ &	 $ 0.1  $\\
		$ r(0) $    &     $   1   $      \\
		$ \boldsymbol{\xi}(0) $	      &   $  [ 0, 0, 0 ]^T $      \\
		$ \boldsymbol{ \sigma}_{EI}(0) $ &  $  [  -0.119, 0.000, 0.159 ]^T  $  \\
		$ \boldsymbol{  \omega}_{EI}^B $	 &  	$[0, 0, 0]^T{\rm rad/s} $ \\
		\bottomrule
	\end{tabular*}
\end{table}

\subsection{Performance of the proposed control scheme }

\begin{figure}[!h]
	\centering
	\begin{minipage}[t]{0.48\textwidth}
		\centering
		\includegraphics[width=0.98\textwidth]{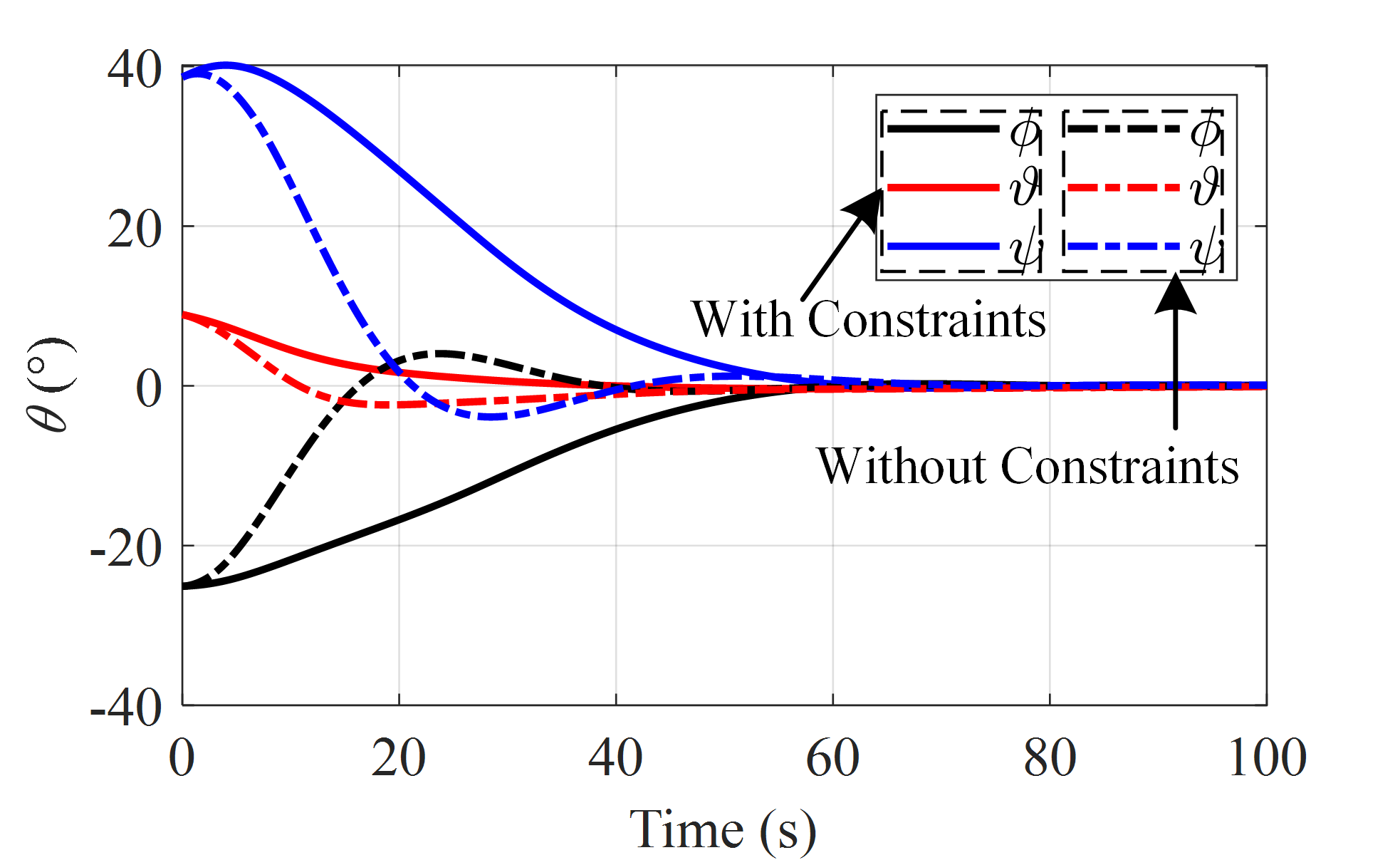}
		\caption{Attitude trajectories.}\label{Attitude}
	\end{minipage}
	\begin{minipage}[t]{0.48\textwidth}
		\centering
		\includegraphics[width=0.98\textwidth]{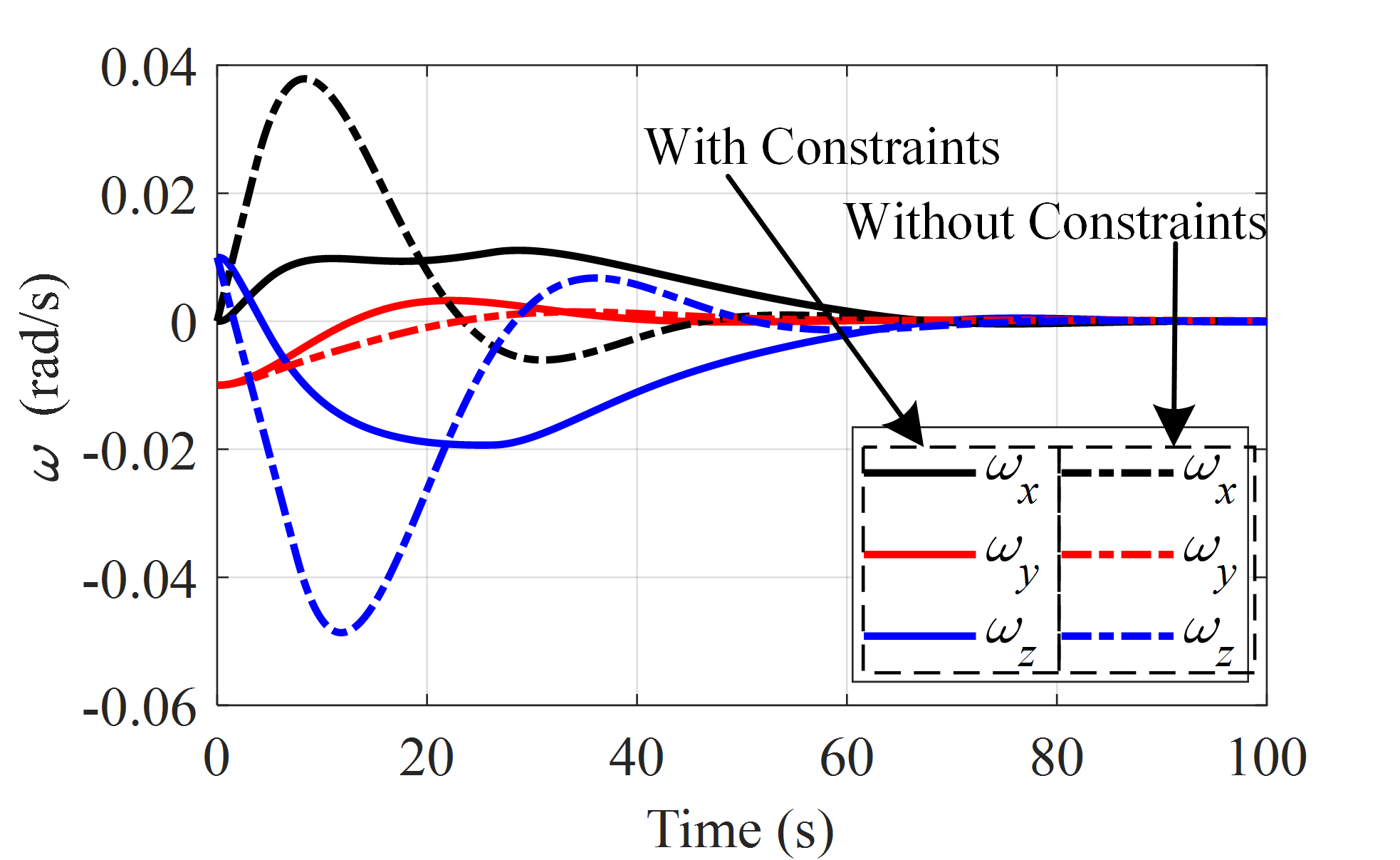}
		\caption{Angular velocities.}\label{A_v}
	\end{minipage}
\end{figure}

\begin{figure}[!h]
	\centering
	\begin{minipage}[t]{0.48\textwidth}
		\centering
		\includegraphics[width=0.98\textwidth]{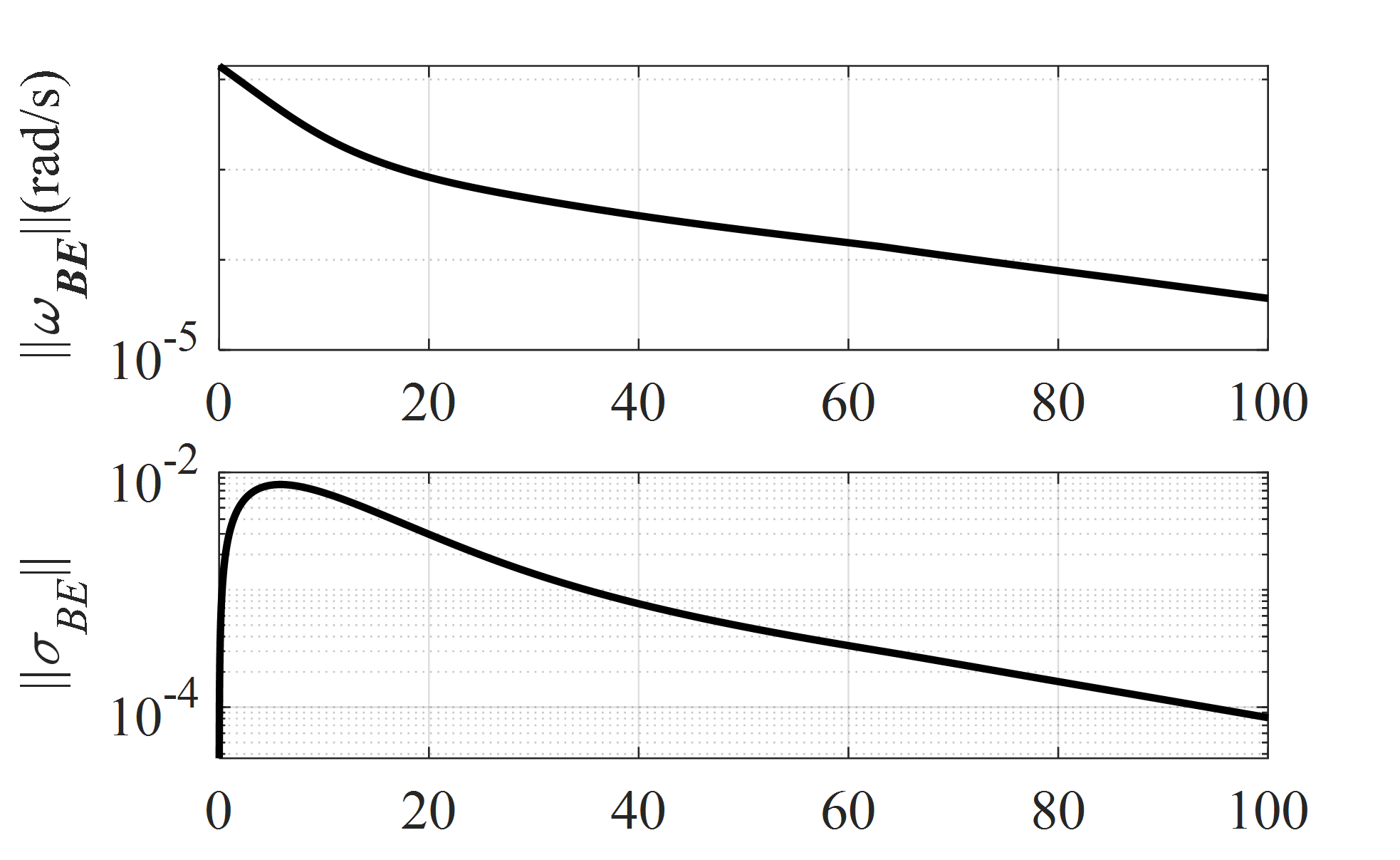}
		\caption{ Attitude  and velocity  estimation errors.}\label{A_v_e}
	\end{minipage}
	\begin{minipage}[t]{0.48\textwidth}
		\centering
		\includegraphics[width=0.98\textwidth]{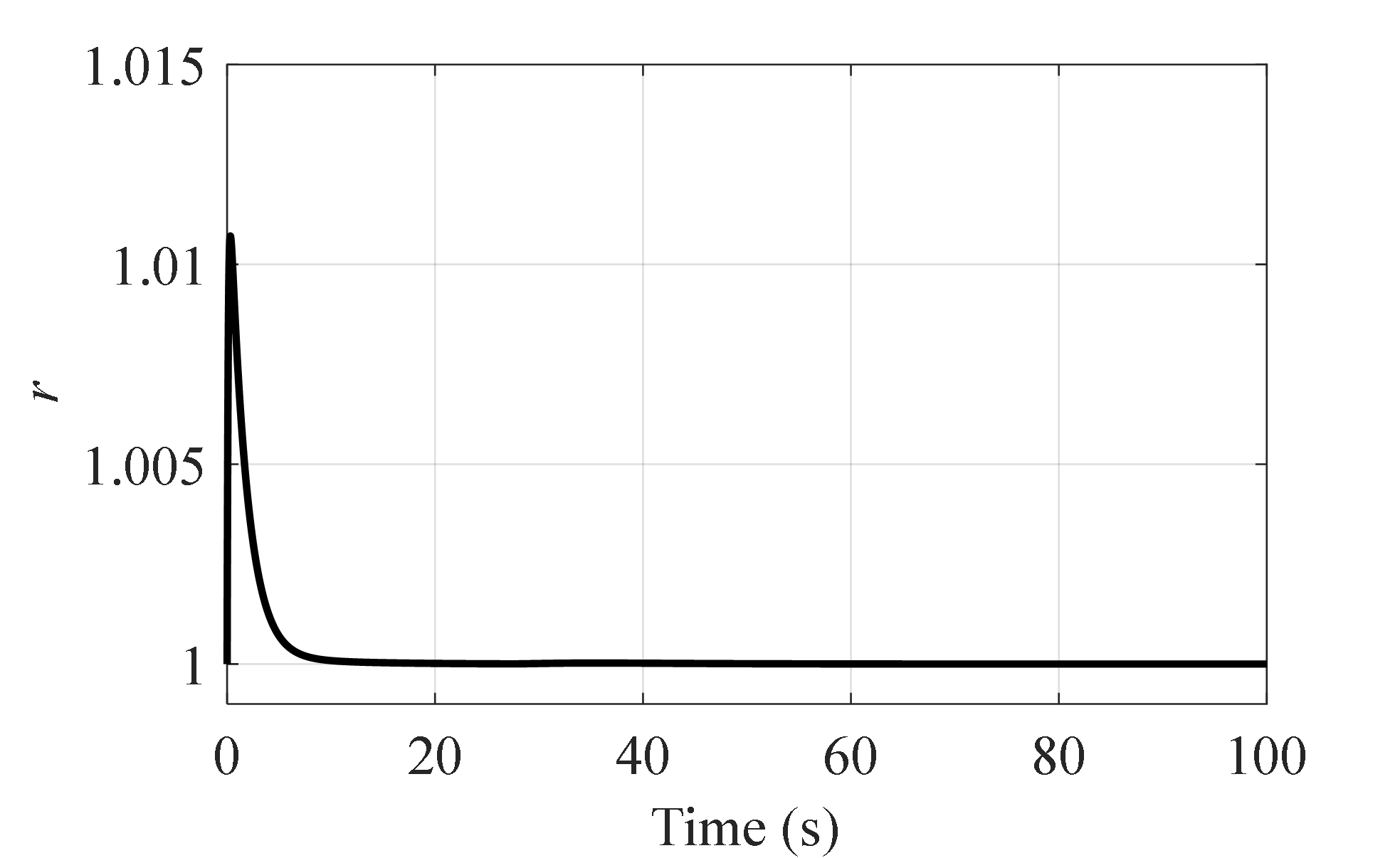}
		\caption{Injection gain r.}\label{r}
	\end{minipage}
\end{figure}

\begin{figure}[!h]
	\centering
	\begin{minipage}[t]{0.48\textwidth}
		\centering
		\includegraphics[width=0.98\textwidth]{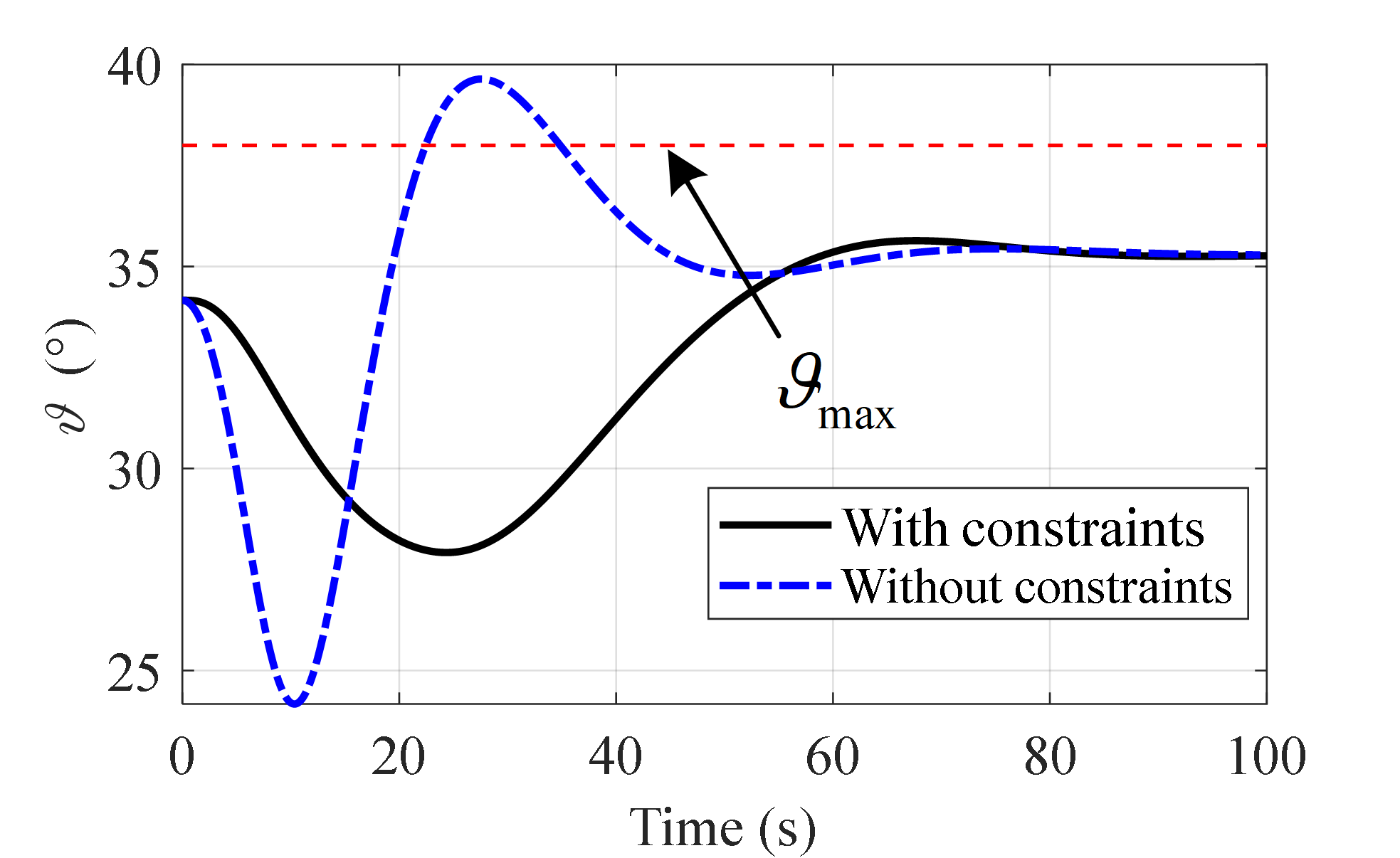}
		\caption{Pointing constraint.}\label{Pointing}
	\end{minipage}
	\begin{minipage}[t]{0.48\textwidth}
		\centering
		\includegraphics[width=0.98\textwidth]{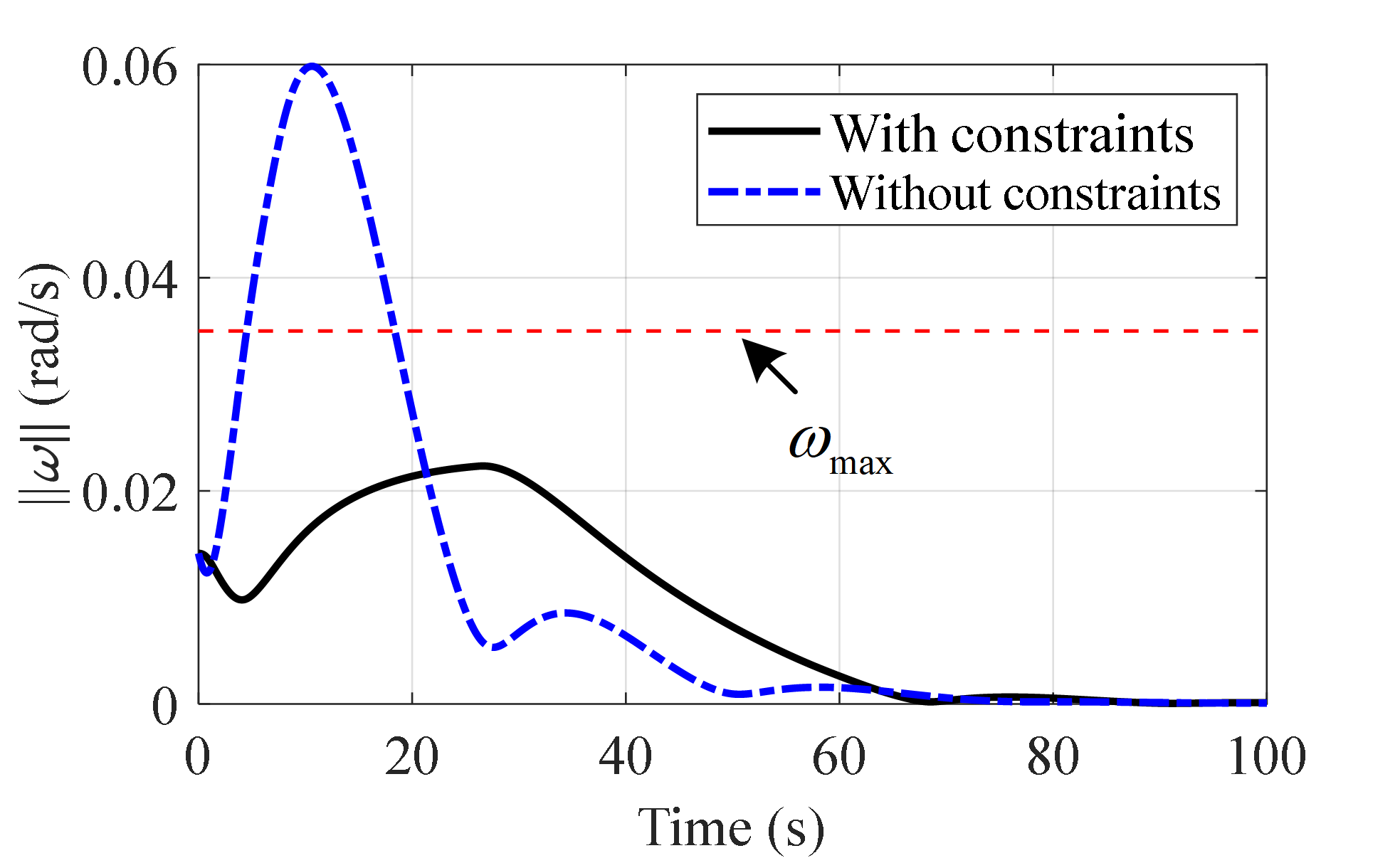}
		\caption{Angular velocity constraint.}\label{W_c}
	\end{minipage}
\end{figure}

\begin{figure}[h]
	\centering
	\begin{minipage}[t]{0.48\textwidth}
		\centering
		\includegraphics[width=0.98\textwidth]{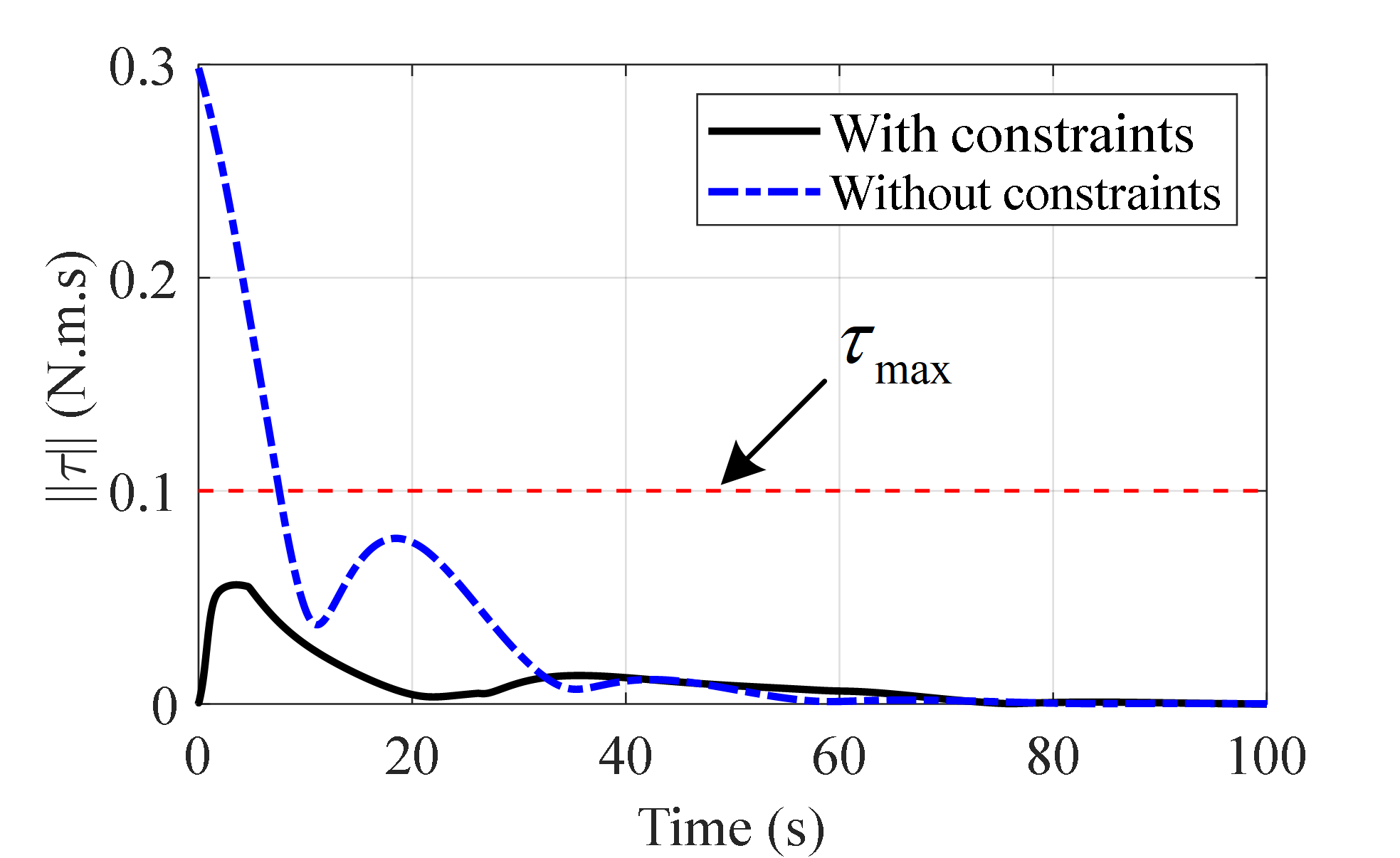}
		\caption{Control torque limitation.}\label{Torque}
	\end{minipage}
	\begin{minipage}[t]{0.48\textwidth}
		\centering
		\includegraphics[width=0.98\textwidth]{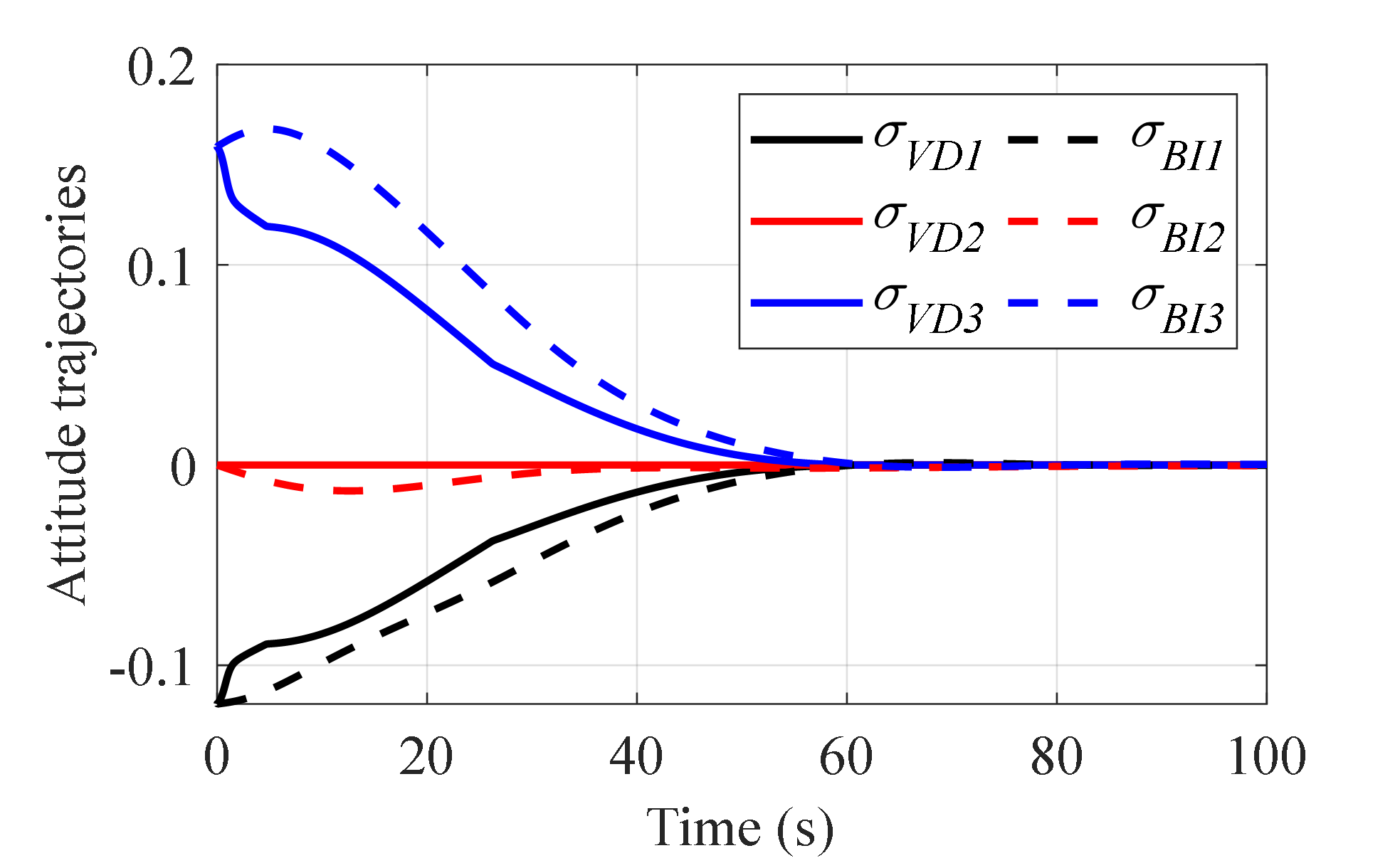}
		\caption{Reference trajectories.}\label{A_t}
	\end{minipage}
\end{figure}

\begin{figure}[h]
	\centering
	\begin{minipage}[t]{0.48\textwidth}
		\centering
		\includegraphics[width=0.98\textwidth]{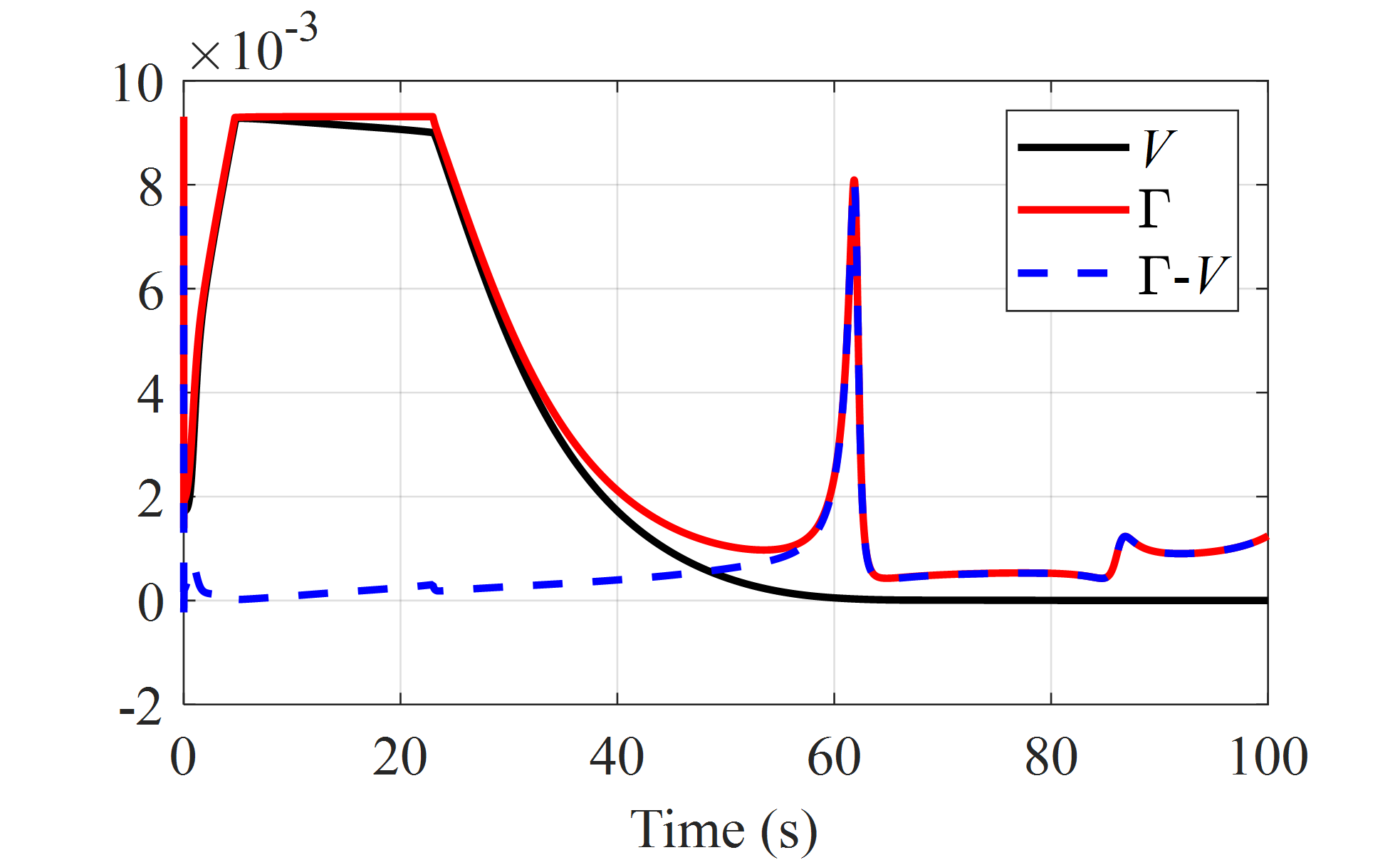}
		\caption{Threshold value.}\label{Threshold}
	\end{minipage}
	\begin{minipage}[t]{0.48\textwidth}
		\centering
		\includegraphics[width=0.98\textwidth]{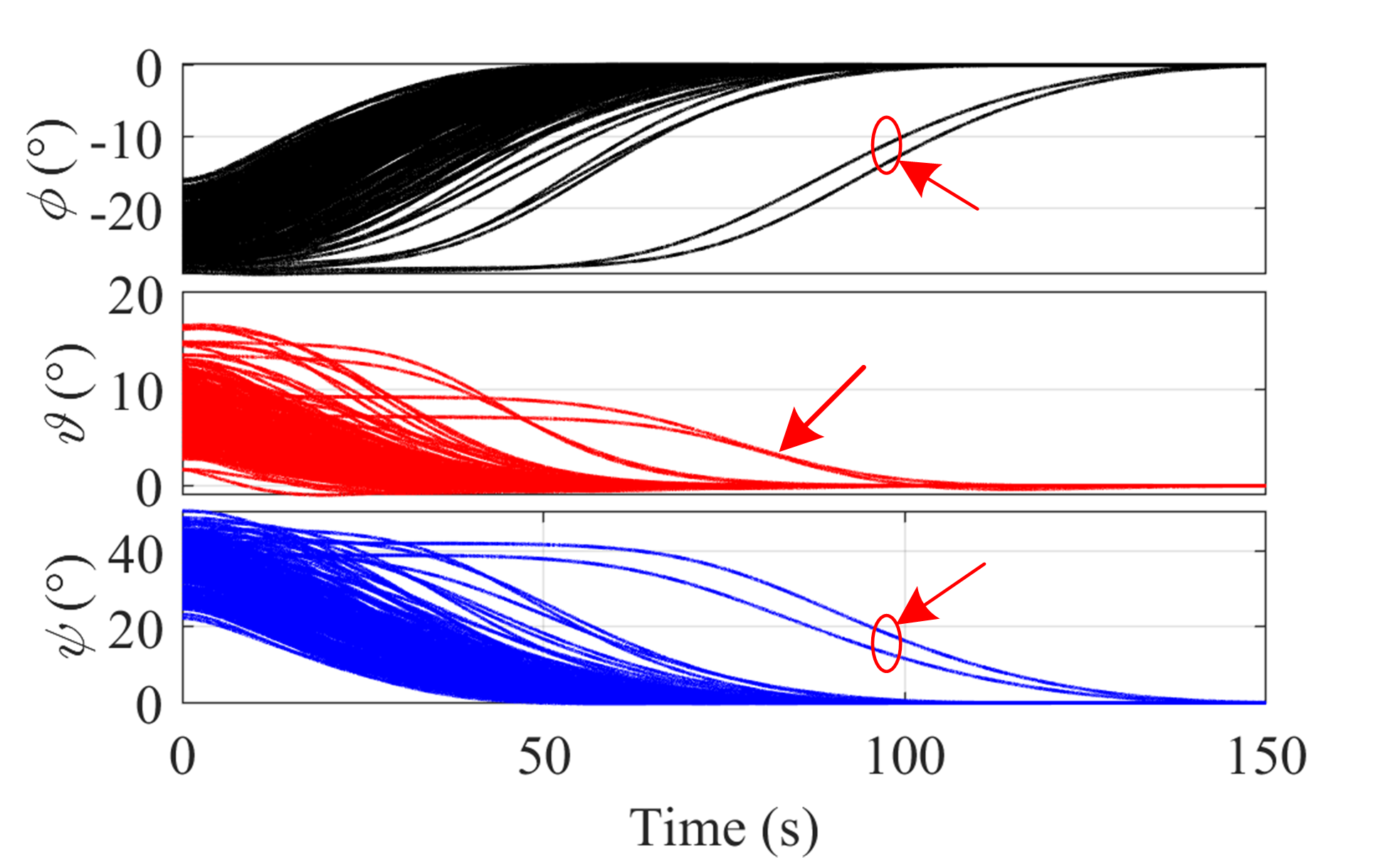}
		\caption{Attitude trajectories.}\label{M_attitude}
	\end{minipage}
\end{figure}

The simulation results are depicted in  Figs. \ref{Attitude} $ - $  \ref{Threshold}, where the dash curves illustrate the simulations  conducted  without reference management i.e.,  the constraints are not  integrated into the control scheme. The  attitude trajectories and angular velocities  depicted in  Figs. \ref{Attitude} and \ref{A_v} indicate that with the navigation layer, the   trajectories of attitude and the velocity become  smoother and the overshoot is smaller.

Fig. \ref{A_v_e} illustrates  the attitude and angular velocity estimation errors generated by the I$ \& $I based observer designed in \eqref{W_O}$-$\eqref{obser_gain}.  In the logarithmic scale, the estimation errors $ \|\boldsymbol{  \omega}_{BE}^B\| $  and $ 	\| \boldsymbol{ \sigma}_{BE} \|$ decreases in an almost straight line, indicating that the estimation errors are exponentially convergent.
An intriguing observation arises:  unlike $ \|\boldsymbol{  \omega}_{BE}^B\| $ decreases consistently over time, $ 	\| \boldsymbol{ \sigma}_{BE} \|$ initially remains very small but undergoes an initial increase before subsequently decreasing.  This behavior arises because  $ 	\| \boldsymbol{ \sigma}_{BE} \|$ is utilized as an "indicator" to assess the adequacy of the estimation of $\| \boldsymbol{  \omega}_{BE}^B\| $. As we set $ 	\| \boldsymbol{ \sigma}_{BE} \|=0$ as the initial condition, and   the estimation error $ \|\boldsymbol{  \omega}_{BE}^B(0)\| $ is substantial, $ 	\| \boldsymbol{ \sigma}_{BE} \|$ grows initially. As  $ \|\boldsymbol{  \omega}_{BE}^B\| $ diminishes, $ 	\| \boldsymbol{ \sigma}_{BE} \|$ subsequently adjusts accordingly.
As we can also seen from Figs. \ref{A_v_e} and \ref{r}, due to the large estimation error at the beginning, the injection gain $ r $ is also relatively large, but as the estimation error diminishes, $ r  $ swiftly converges towards 1.
These outcomes underscore the pivotal role of the injection gain $ r $ in the observer system, contributing significantly to achieving the desired effectiveness of the observer.
Furthermore, the final angular velocity estimation error $ \|\boldsymbol{  \omega}_{BE}^B(0)\| $ measures approximately $ 10^{-4} $, which is mainly restricted by the simulation setting 0.01s. If the step size is further reduced, $ \|\boldsymbol{  \omega}_{BE}^B(0)\| $ can also be reduced.

The pointing constraint, angular velocity constraint and the control torque limitation are plotted in  Figs. \ref{Pointing} $ -$ \ref{Torque}. Evidently, in the absence of  reference management, the pointing angle exceeds   the boundary at 23 seconds,   the angular velocity surpasses the limitation around 15 seconds, and the actuator saturation occurs within the initial 10 seconds.
Upon the implementation of the navigation layer within the system, all three constraints remain within permissible bounds. 
This is because the tracking error can always maintains  a   small error relative to the reference trajectories (see Fig. \ref{A_t}). 
Analyzing Figs. \ref{A_t} and \ref{Threshold}, reveals that the convergence speed of the reference trajectory aligns closely with the threshold error $  \Gamma- V $.
Overall, the simulation results are in line with the theoretical analysis, verifying the effectiveness of the algorithm.

\subsection{Monte Carlo simulation under disturbances}

The aforementioned section  presented   numerical simulations without any disturbance i.e., $ \boldsymbol{\tau}_d=\textbf{0}_{3\times1} $.
This section conducts Monte Carlo simulations encompassing disturbances to showcase the robustness of the proposed ERG and I\&I technology-based attitude control scheme.
To conduct the Monte Carlo simulations, randomized initial conditions and parameters are detailed in Table  \ref{table3}. These newly introduced initial states and parameters are integrated with previously selected simulation conditions mentioned in the previous section, amounting to a total of 200 Monte Carlo simulations.
In order to ensure that all the cases can reach the final states,
those cases that do not meet the constraints at the initial moment will be excluded, and the simulations last  for 150 seconds.  Besides, the external disturbances are given as follows

\begin{equation*}
	\boldsymbol{\tau}_d^{I}=\begin{pmatrix}\begin{bmatrix}2\\-1\\-3\end{bmatrix}
		+\begin{bmatrix}0.4\mathrm{sin}(\omega_At+1.6)\\2\mathrm{sin}(\omega_At+1.1)\\0.7\mathrm{sin}(\omega_At-2.1)\end{bmatrix}
	\end{pmatrix}
	\times10^{-5}\mathrm{N.m}^{2}\nonumber
\end{equation*}
where $ \omega_{At}=0.01 rad/s $.

\begin{figure}[ht]
	\centering
	\begin{minipage}[t]{0.485\textwidth}
		\centering
		\includegraphics[width=0.95\textwidth]{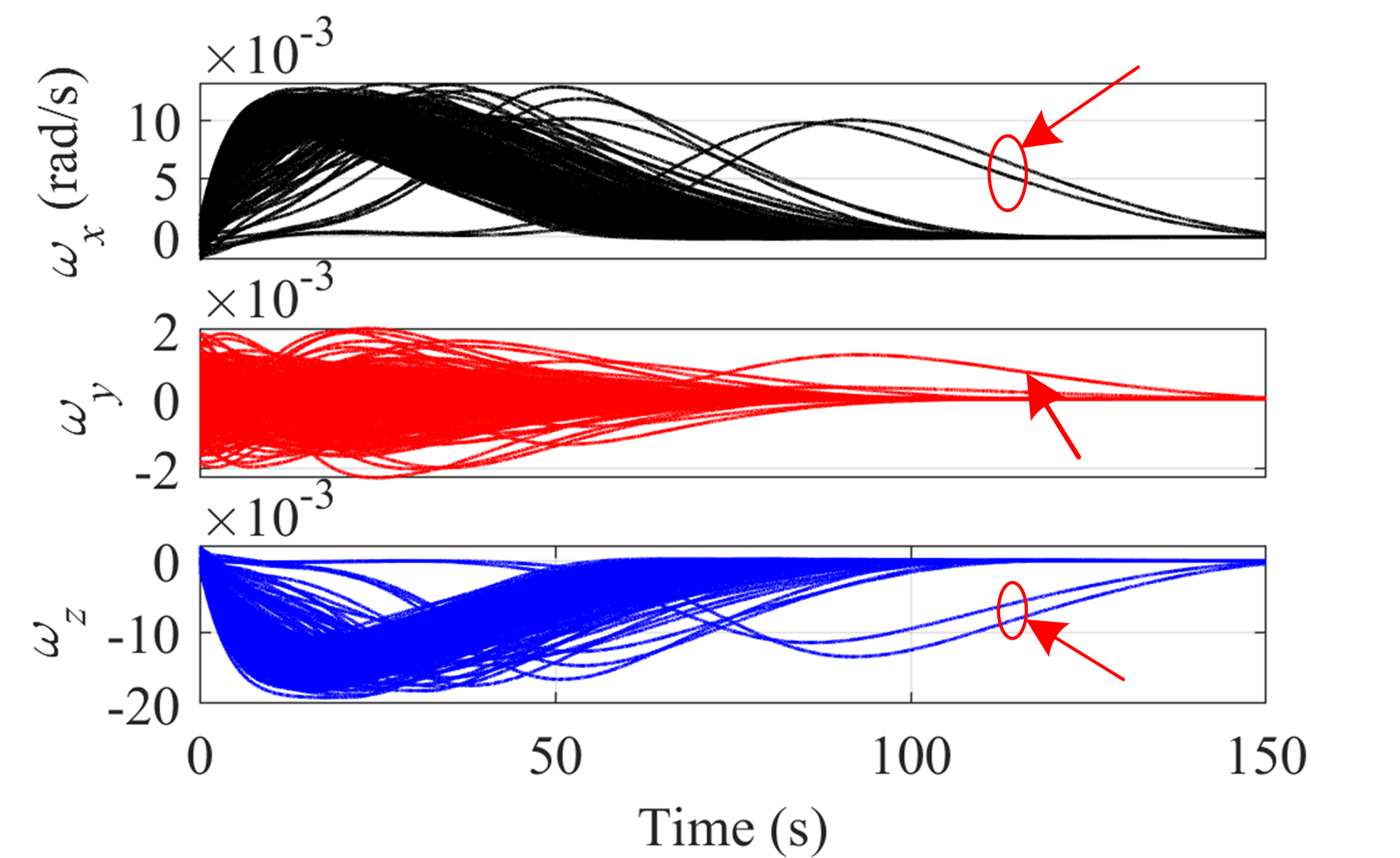}
		\caption{Angular velocity  trajectories.}\label{M_wega}
	\end{minipage}
	\begin{minipage}[t]{0.495\textwidth}
		\centering
		\includegraphics[width=0.95\textwidth]{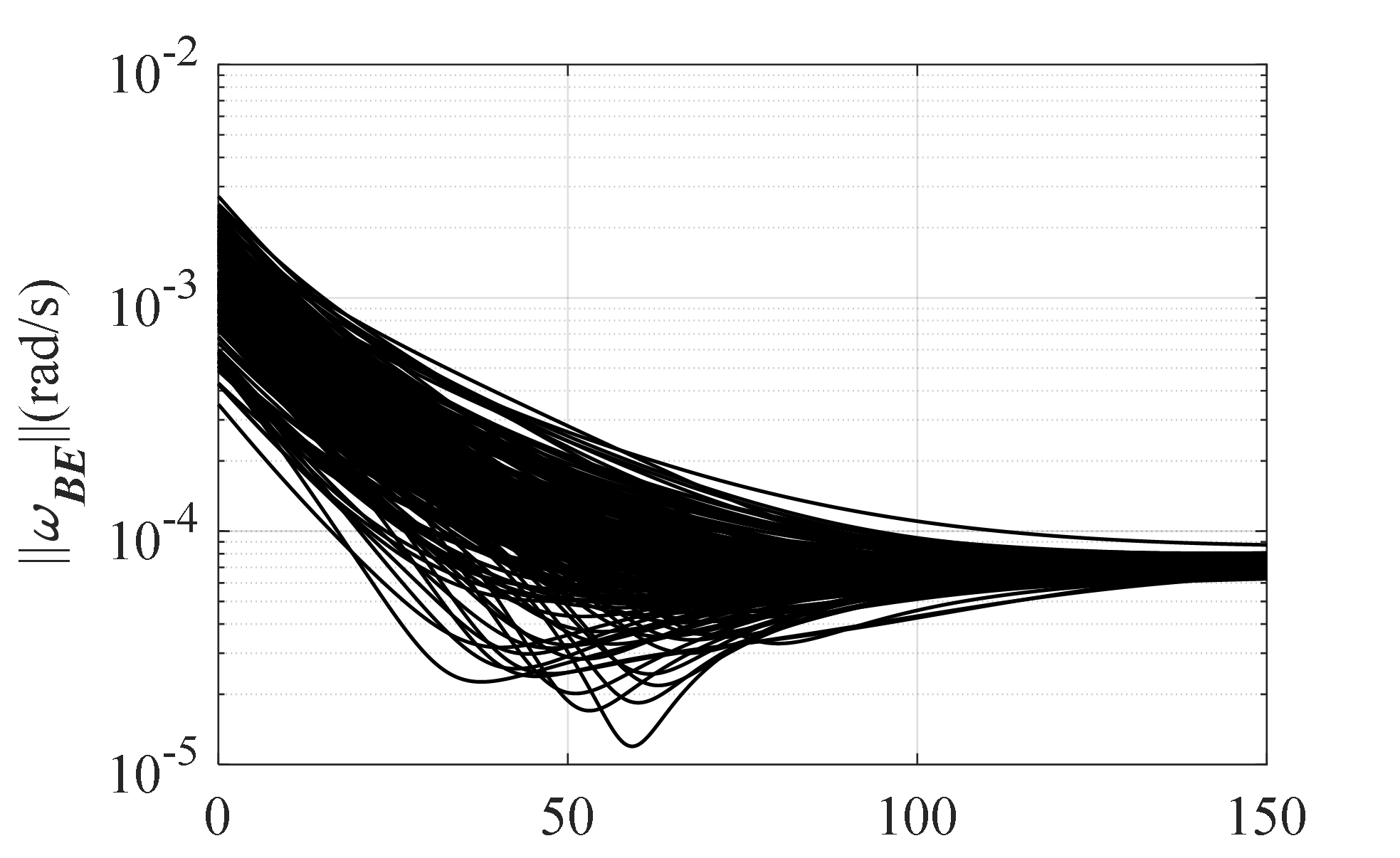}
		\caption{Angular velocity  estimation errors. }\label{M_W_be}
	\end{minipage}
\end{figure}

\begin{figure}[ht]
	\centering
	\begin{minipage}[t]{0.485\textwidth}
		\centering
		\includegraphics[width=0.95\textwidth]{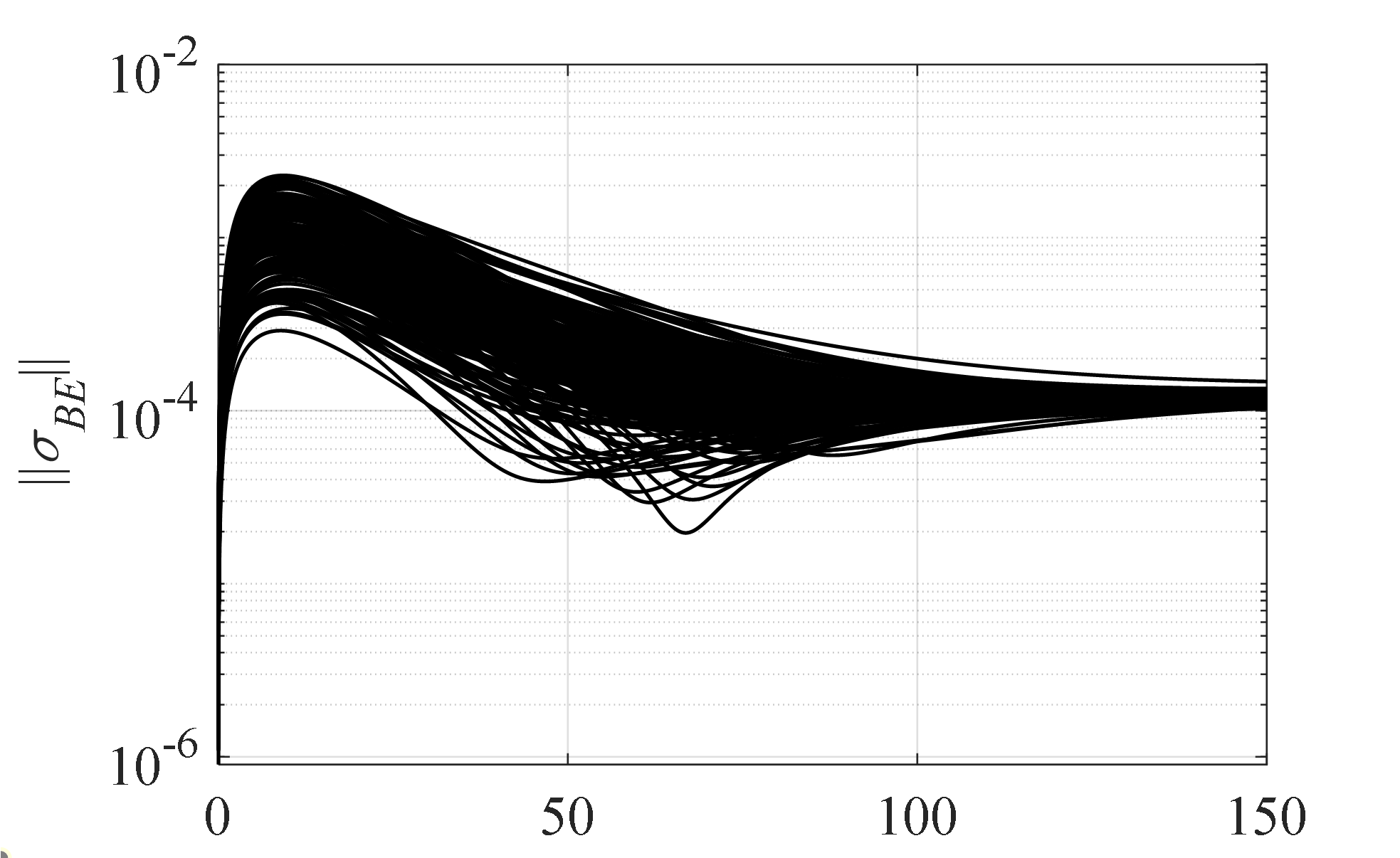}
		\caption{Attitude estimation errors.}\label{M_a_be}
	\end{minipage}
	\begin{minipage}[t]{0.495\textwidth}
		\centering
		\includegraphics[width=0.95\textwidth]{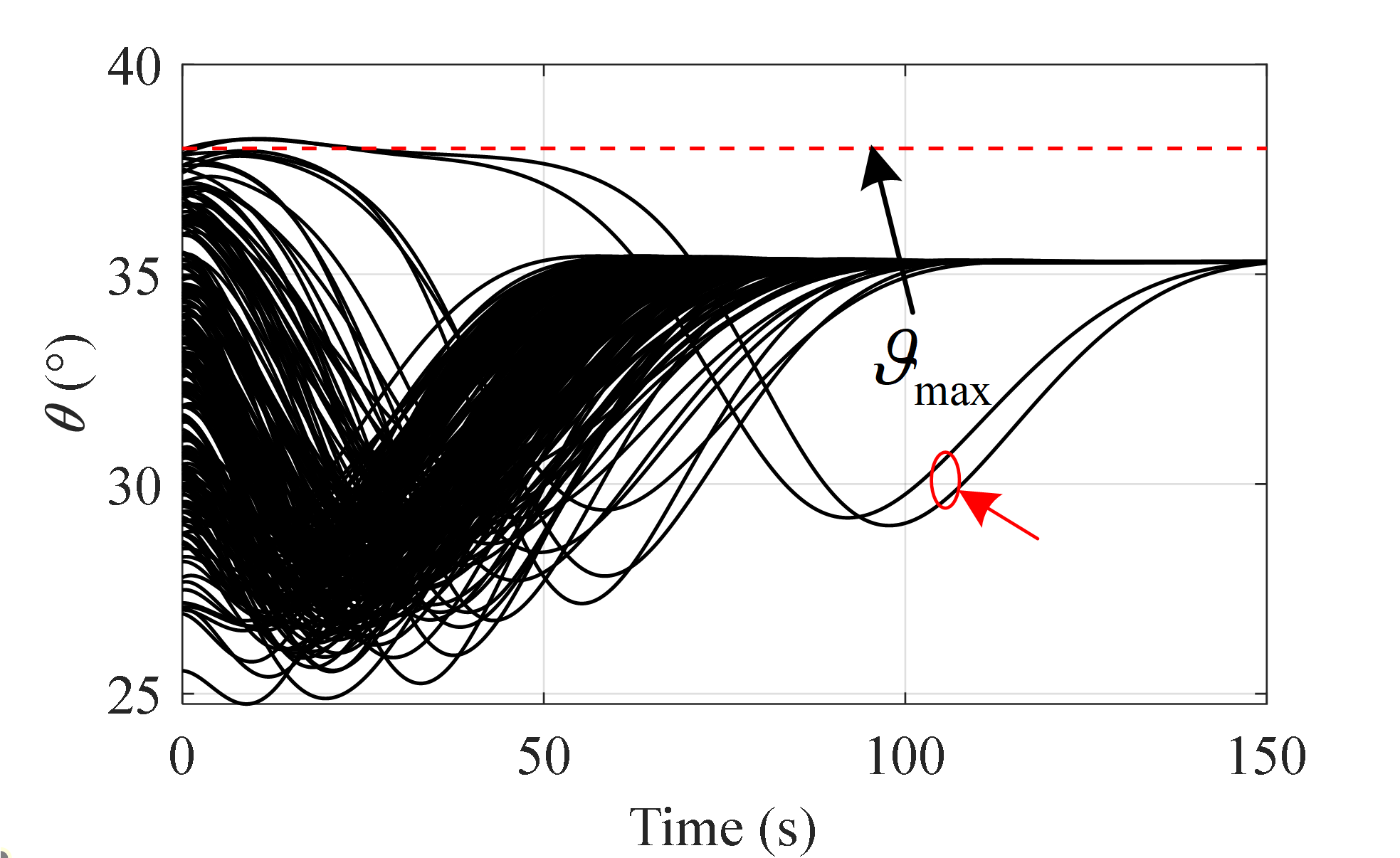}
		\caption{Distributions of pointing angles.}\label{M_a}
	\end{minipage}
\end{figure}

\begin{figure}[ht]
	\centering
	\begin{minipage}[t]{0.485\textwidth}
		\centering
		\includegraphics[width=0.95\textwidth]{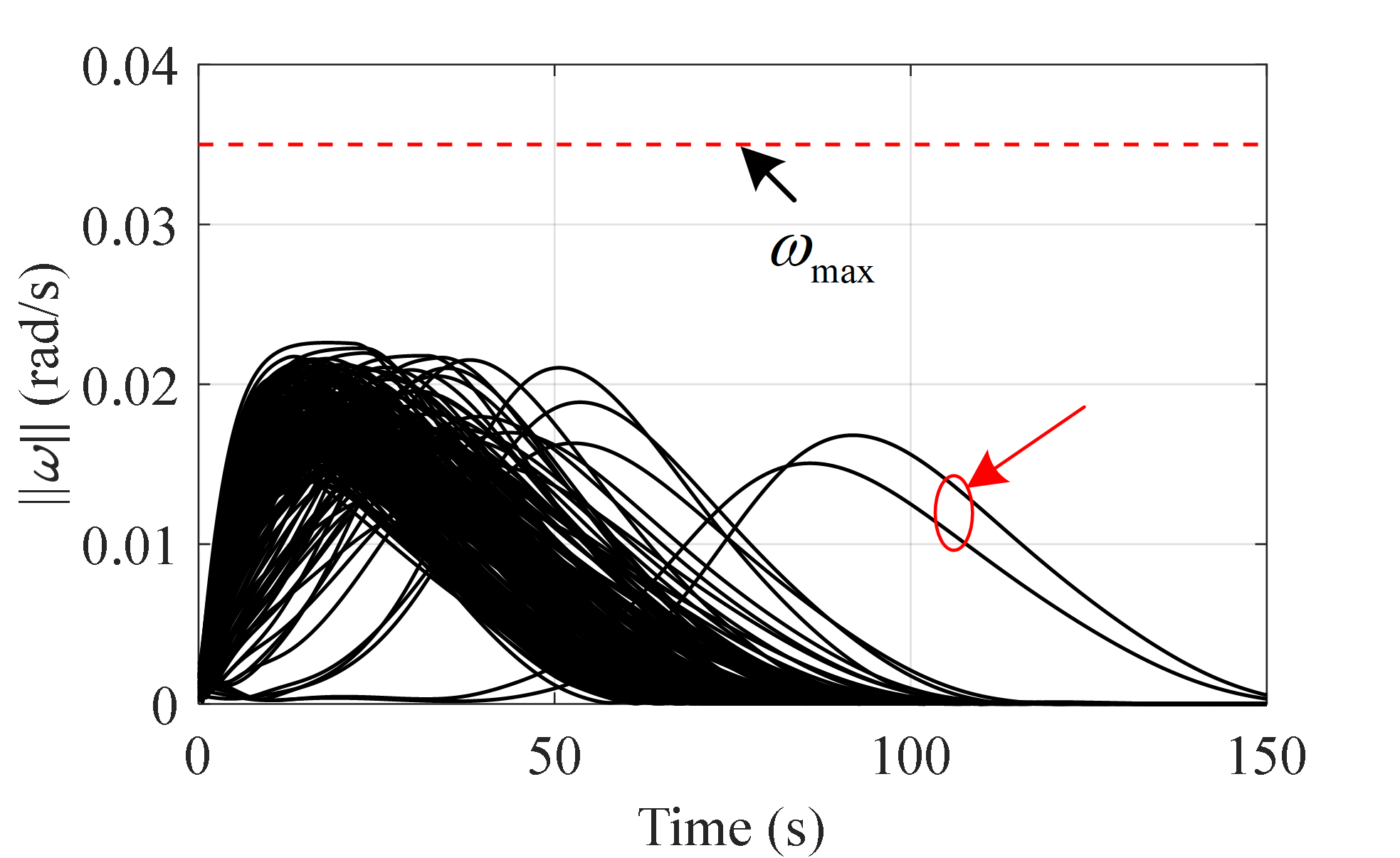}
		\caption{Distributions of angular velocity constraints.}\label{M_w}
	\end{minipage}
	\begin{minipage}[t]{0.495\textwidth}
		\centering
		\includegraphics[width=0.95\textwidth]{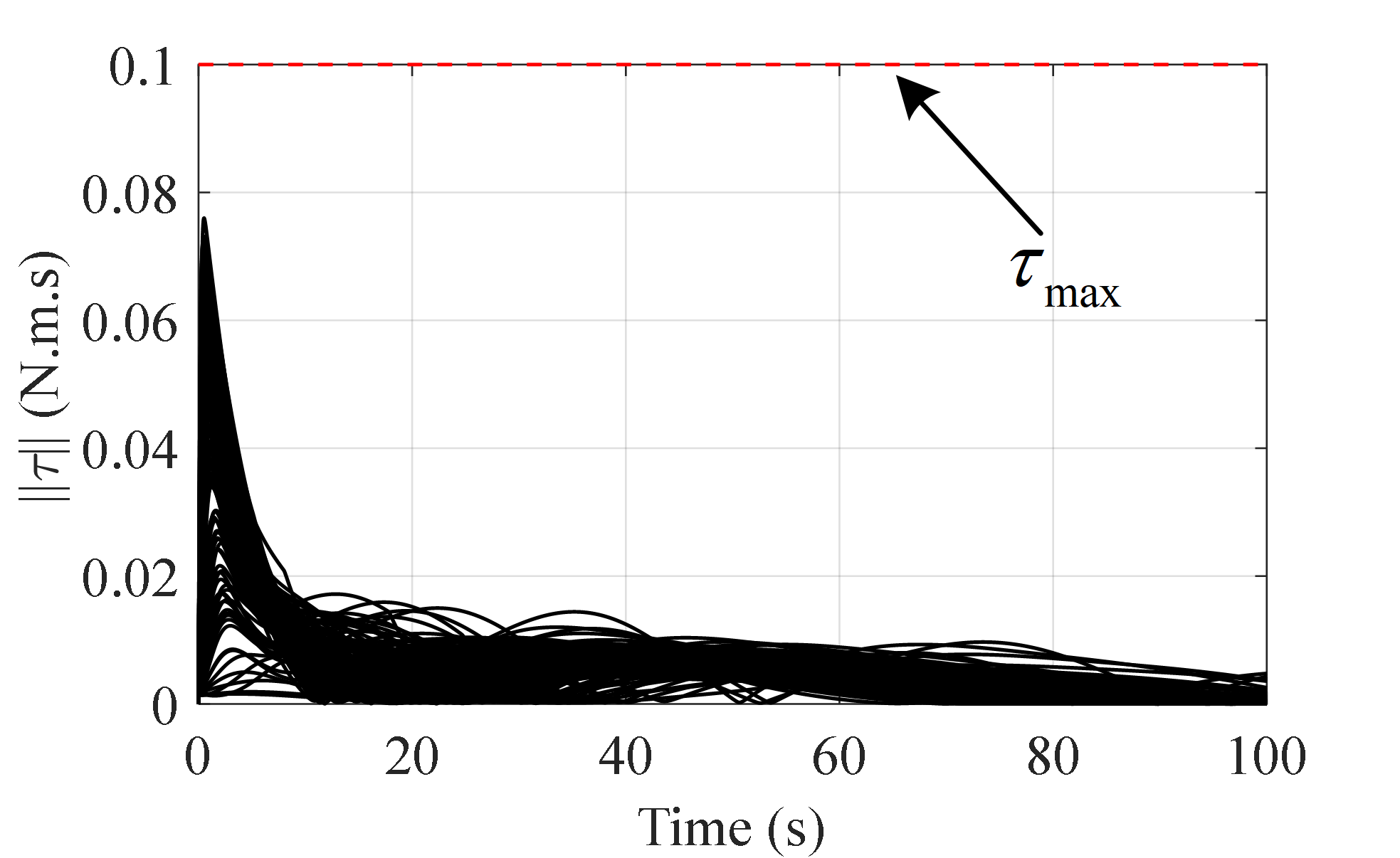}
		\caption{Distributions of control torque.}\label{M_t}
	\end{minipage}
\end{figure}

Figs. \ref{M_attitude} and \ref{M_wega} depict the attitude and angular velocity trajectories of  Monte Carlo simulations. Despite external disturbances and angular velocity estimation errors influencing the control performance, the spacecraft achieved the desired state smoothly.
It can be seen from   \eqref{W_O} and \eqref{W_O_detail} that the precision of the input torque significantly influences estimation errors.  
In the Monte Carlo simulations, the presence of unknown disturbances results in slightly larger estimation errors, as depicted in Figs. \ref{M_W_be} and \ref{M_a_be}, compared to those in Fig. \ref{A_v_e}. Nevertheless, these errors remain comparatively small, affirming the observer's robustness as designed in \eqref{W_O}.
Besides, the three constraints are basically satisfied during the simulation (see  Figs. \ref{M_a} $ - $ \ref{M_t}).
Meanwhile, one can also see that the angular velocity observer is independent of the   controller. No matter what constraints the system needs to meet, the observer can converge without being affected by them.  This intriguing property enables us to enhance the ERG-based constrained controller without limitations imposed by the observer.

\begin{center}
	\begin{table}[h]%
		\centering
		\caption{Randomized initial states and parameters.\label{table3}}%
		\begin{tabular*}{200pt}{@{\extracolsep\fill}lcccc@{\extracolsep\fill}}
			\toprule
			\textbf{	Variables} & \textbf{Ranges}  \\
			\midrule
			$ 	 \textbf{\textit{n}} (0)  $      &  $[-0.7,-0.5] \times [-0.1,0.1] \times[0.7,0.9]  $  \\
			$ \phi (0)  $, rad &  $ [0.15\pi, 0.35\pi] $  \\
			$\boldsymbol{  \omega}(0)$, rad/s      &$\{ [-2,2] \times [-2,2] \times[-2,2]\} \times10^{-3} $  \\
			$ J_m $  &   [13, 17]\\
			$ J_M $  &   [15, 21]\\
			$ k_p=1.5 $  &   [1, 1.5]\\	
			$  k_d=2.5 $ &   [2.5, 3.0]\\
			$ k_e=1000 $&  [900, 1100] \\
			\bottomrule
		\end{tabular*}
	\end{table}
\end{center}

However, a slight setback is identified as there are two cases of slow convergence (as denoted by the red circles in Figs. \ref{M_attitude}, \ref{M_wega}, \ref{M_a}, and \ref{M_w}). This occurrence is due to these specific cases being at the verge of the pointing constraint initially.
Due to external disturbances and angular velocity estimation errors, the pointing angle slightly overflows the boundary.
Fortunately, the algorithm  designed in this study addresses these effects by employing an enhanced safety margin as described in \eqref{deta}, swiftly reeling back the pointing angle. 
To prevent such occurrences rigorously, adjusting the reference state's margin suffices, a topic to be extensively explored in our subsequent research.
In spite of this, the proposed ERG-based constrained  controller still accomplished the maneuver objective with remarkable  accuracy, which in turn demonstrates its robustness against uncertainties.

\section{Conclusion}

This paper develop  a   constrained output feedback attitude reorientation problem via ERG and I$ \& $I technologies, where pointing constraint, angular velocity constraint, and the control saturation are considered.
The stability  of a angular velocity observer and the output feedback controller is roughly proved.
The inner loop of ERG is conducted by the angular velocity observer-based output feedback controller, and the navigation layer is designed by manipulating the auxiliary reference state without violating the constraints while asymptotically converges to the desired reference.
The performance of the proposed angular velocity observer and the ERG is meticulously analyzed and discussed by numerical simulations in detail.

\section*{Acknowledgments}
This work was partially supported by National Natural Science Foundation of China under grants 62273240, the Natural Science Foundation of Shaanxi Province (Grant No. 2023-JC-QN-0003),  the Fundamental Research Funds for the Central Universities(Grant No.  23GH020210),  Suzhou Municipal Science and Technology Bureau under Grant (ZXL2023177), and the basic Research Program of Taicang, Grant (TC2023JC13).

\section*{Conflict of interest}

The authors declare no potential conflict of interests.

\section*{Declaration of generative AI and AI-assisted technologies in the writing process}

During the preparation of this work the authors used  ChatGPT  in order to  improve language and readability. After using this tool, the authors reviewed and edited the content as needed and take full responsibility for the content of the publication.

\appendix

\section{Proof of proposition1\label{app1}}

\textbf{Proof}:
The  \textbf{proposition1} can be proven by Lyapunov method by two steps. This first step is prestabilizing the inject estimation error $ \textbf{\textit{z}} $. The second step is ensuring the dynamic scaling  $ r $ is bounded.

The Lyapunov function candidate about the  $ \textbf{\textit{z}} $ is chosen as
\begin{equation}
	V_z=\frac{1}{2}\textbf{\textit{z}}^T\textbf{\textit{J}}\textbf{\textit{z}}.
\end{equation}
The dynamics of $  \boldsymbol{  \omega}_{EI}^B $ can be obtained from \eqref{W_O}, \eqref{W_O1},  \eqref{segma_err} that
\begin{equation}\label{dyn_WE}
	\dot{\boldsymbol{  \omega}}_{EI}^B=\textbf{\textit{J}}^{-1}(-\boldsymbol{  \omega}_{EI}^B\times\textbf{\textit{J}}\boldsymbol{  \omega}_{EI}^B+\boldsymbol{\tau}_c^{B})+
	4\textbf{\textit{J}}^{-1}\beta(\underline{\varpi } )\dot{\boldsymbol{ \sigma}}_{2BE}.
\end{equation}
One can derive  the dynamics of $ \boldsymbol{  \omega}_{BE}^B $ from   \eqref{Dyn_2}, \eqref{W_error} , and \eqref{dyn_WE} that
\begin{equation} \label{W_BE}
	\dot{\boldsymbol{\omega}}_{BE}^B=\textbf{\textit{J}}^{-1}( \boldsymbol{  \omega}_{EI}^B\times\textbf{\textit{J}}\boldsymbol{  \omega}_{EI}^B -\boldsymbol{  \omega}_{BI}^B\times\textbf{\textit{J}}\boldsymbol{  \omega}_{BI}^B )-
	4\textbf{\textit{J}}^{-1}\beta(\underline{\varpi } )\dot{\boldsymbol{ \sigma}}_{2BE}.
\end{equation}	
Then, by invoking   \eqref{re_2}, \eqref{r_dyn} \eqref{obser_gain1},  and \eqref{obser_gain2}   and using the  inequalities $  \|\boldsymbol{  \omega}_{EI}^B\|\leq\varpi $, $ \|\textbf{\textit{a}}\| \leq \| \textbf{\textit{a}}-\textbf{\textit{b}}\|+\|\textbf{\textit{b}}\| $, and $ 2\textbf{\textit{a}}\textbf{\textit{b}}\leq \|\textbf{\textit{a}}\|^2 +\|\textbf{\textit{b}}\|^2$,   the  time derivative of $ V_z $ along \eqref{z} and \eqref{W_BE} can be obtained as
\begin{equation}\label{d_Vz}
	\begin{split}
		\dot V_z=&\textbf{\textit{z}}^T\textbf{\textit{J}}	r^{-1}
		\textbf{\textit{J}}^{-1}( \boldsymbol{  \omega}_{EI}^B\times\textbf{\textit{J}}\boldsymbol{  \omega}_{EI}^B -\boldsymbol{  \omega}_{BI}^B\times\textbf{\textit{J}}\boldsymbol{  \omega}_{BI}^B ) \\
		&-\textbf{\textit{z}}^T\textbf{\textit{J}}\left\lbrace r^{-1}4\textbf{\textit{J}}^{-1}\beta(\underline{\varpi } )\dot{\boldsymbol{ \sigma}}_{2BE}
		-\dot r^{-1}\boldsymbol{\omega}_{BE}^B
		\right\rbrace \\
		\leq&\textbf{\textit{z}}^T\left(  \boldsymbol{  \omega}_{EI}^B\times\textbf{\textit{J}}\textbf{\textit{z}}
		-\underline{\beta}(\underline{\varpi } )\textbf{\textit{z}}\right) - r^{-1}\textbf{\textit{z}}^T\textbf{\textit{J}}\dot r
		\textbf{\textit{z}}\\
		\leq&J_M \| \underline{\varpi } \|\| \textbf{\textit{z}} \|^2-\underline{\beta}(\underline{\varpi } )\| \textbf{\textit{z}} \|^2+\dfrac{J_mk_r}{J_M}\| \textbf{\textit{z}} \|^2\\
		=&-(1+\rho_{\varpi })\| \textbf{\textit{z}} \|^2
	\end{split}
\end{equation}
which implies $ \textbf{\textit{z}} $ converges to zeros exponentially.

To show the boundedness of $ \textbf{\textit{r}} $, consider
\begin{equation}
	V_o=V_z+V_\varpi+V_\sigma+V_r
\end{equation}
where
\begin{equation}
	V_\varpi=\frac{1}{2}(\varpi-\underline\varpi)^2\nonumber
\end{equation}
\begin{equation}
	V_\sigma=2{\rm ln}( 1+\sigma^2_{BE})\nonumber
\end{equation}
\begin{equation}
	V_r= \dfrac{J_m}{2}(r-1)^2\nonumber.
\end{equation}
Using \eqref{W_Cover} and \eqref{dyn_WE}, the dynamics of $ \varpi $ is derived as
\begin{equation}\label{W_cover_e}
	\begin{aligned}
		\dot\varpi=&\varpi^{-1}(\boldsymbol{  \omega}_{EI}^B )^T\dot{ \boldsymbol{  \omega}}_{EI}^B\\
		=&\varpi^{-1}(\boldsymbol{  \omega}_{EI}^B )^T	\textbf{\textit{J}}^{-1}\!\left\lbrace\!
		-\boldsymbol{\omega}_{EI}^B\!\times\!\textbf{\textit{J}}\boldsymbol{  \omega}_{EI}^B+\boldsymbol{\tau}_c^{B}+
		4\beta(\underline{\varpi } )\dot{\boldsymbol{ \sigma}}_{2BE}
		\right\rbrace\!.
	\end{aligned}
\end{equation}
From   \eqref{W_O2} and \eqref{W_cover_e}, it follows that
\begin{equation} \label{W_error_E}
	\dot \varpi- \dot {\underline\varpi}=\varpi^{-1}(\boldsymbol{  \omega}_{EI}^B )^T
	4\textbf{\textit{J}}^{-1}\beta(\underline{\varpi } )\dot{\boldsymbol{ \sigma}}_{2BE}
	-K_{\underline{\varpi }}(\varpi-\underline\varpi)
\end{equation}
Take the time derivative of $ V_\varpi $ along \eqref{W_error_E}, one can obtain
\begin{equation}\label{d_W}
	\begin{aligned}
		\dot V_\varpi=&(\varpi\!-\!\underline\varpi)\!\left\lbrace
		\varpi^{-1}(\boldsymbol{  \omega}_{EI}^B )^T
		\!	4\textbf{\textit{J}}^{-1}\beta(\underline{\varpi } )\dot{\boldsymbol{ \sigma}}_{2BE}
		\!-\!K_{\underline{\varpi }}(\varpi\!-\!\underline\varpi)
		\right\rbrace \\
		\leq&\frac{1}{2}\textbf{\textit{z}}^2-(\varpi-\underline\varpi)^2\left\lbrace
		K_{\underline{\varpi }}-8\left(\dfrac{\|\boldsymbol{  \omega}_{EI}^B \|\beta(\underline{\varpi } )r}{J_m\varpi} \right)^2
		\right\rbrace
	\end{aligned}
\end{equation}
Applying   \eqref{segma_err}  and \eqref{re_1}, one has

\begin{equation}\label{d_se_err}
	\begin{aligned}
		\dot V_\sigma=& \dfrac{4\boldsymbol{\sigma}^T_{BE}}{1+\boldsymbol{ \sigma}^2_{BE}}G(\boldsymbol{ \sigma}_{BE})
		( \boldsymbol{  \omega}_{BE}^B-K_\sigma\boldsymbol{ \sigma}_{BE}  )\\
		\leq&\frac{1}{2}\textbf{\textit{z}}^2-\left(K_\sigma-\frac{1}{2}r^2 \right) \|\boldsymbol{ \sigma}_{BE} \|^2
	\end{aligned}
\end{equation}
Additionally, following the calculations in   \eqref{r_dyn},  one can obtain

\begin{equation}\label{d_r}
	\begin{aligned}
		\dot V_r=&  J_m(r-1)\left\lbrace
		\dfrac{\textit{r}}{J_m}(J_M\|  \varpi  -\underline{\varpi } \|)-\dfrac{k_r}{J_M}(r-1)
		\right\rbrace \\
		\leq&\frac{1}{2}\textbf{\textit{r}}^2J_M\|  \varpi  -\underline{\varpi } \|^2-
		\left( \dfrac{J_mk_r}{J_M}-\frac{1}{2}J_M \right) (r-1)^2
	\end{aligned}
\end{equation}

Finally, differentiating $ V_o $ along  \eqref{d_Vz}, \eqref{d_W}, \eqref{d_se_err}, and \eqref{d_r}, yields
\begin{equation}
	\dot V_o\leq - \rho_\varpi\| \textbf{\textit{z}} \|^2-\rho_{ \underline\varpi }(\varpi-\underline\varpi)^2
	-\rho_\sigma\|\boldsymbol{ \sigma}_{BE} \|^2-\rho_r(r-1)^2
\end{equation}
which implies that the system is exponentially stable and $ r $ is bounded. This completes the proof.



\bibliographystyle{elsarticle-num}
\bibliography{mybibfile}

\begin{thebibliography}{10}
\expandafter\ifx\csname url\endcsname\relax
  \def\url#1{\texttt{#1}}\fi
\expandafter\ifx\csname urlprefix\endcsname\relax\def\urlprefix{URL }\fi
\expandafter\ifx\csname href\endcsname\relax
  \def\href#1#2{#2} \def\path#1{#1}\fi

\bibitem{8424433}
C.~Wei, Q.~Chen, J.~Liu, Z.~Yin, J.~Luo, An overview of prescribed performance
  control and its application to spacecraft attitude system, Proceedings of the
  Institution of Mechanical Engineers, Part I: Journal of Systems and Control
  Engineering 235~(4) (2021) 435--447.

\bibitem{QI2023292}
R.~Qi, X.~Dong, D.~Chao, Y.~Wang, Constrained attitude tracking control and
  active sloshing suppression for liquid-filled spacecraft, ISA Transactions
  132 (2023) 292--308.

\bibitem{NAKANO2023111103}
S.~Nakano, T.~W. Nguyen, E.~Garone, T.~Ibuki, M.~Sampei, Explicit reference
  governor on so(3) for torque and pointing constraint management, Automatica
  155 (2023) 111103.

\bibitem{QU202283}
Y.~Qu, X.~Zhong, F.~Zhang, X.~Tong, L.~Fan, L.~Dai, Robust disturbance
  observer-based fast maneuver method for attitude control of optical remote
  sensing satellites, Acta Astronautica 201 (2022) 83--93.

\bibitem{G4469}
H.~B. Hablani, Attitude commands avoiding bright objects and maintaining
  communication with ground station, Journal of Guidance, Control, and Dynamics
  22~(6) (1999) 759--767.

\bibitem{CS}
H.~Cui, X.~Cheng, Anti-unwinding attitude maneuver control of spacecraft
  considering bounded disturbance and input saturation, SCIENCE CHINA
  Technological Sciences 55~(9) (2012) 2518--2529.

\bibitem{DANG2021}
Q.~Dang, K.~Liu, J.~Wei, Explicit reference governor based spacecraft attitude
  reorientation control with constraints and disturbances, Acta Astronautica
  190 (2022) 455--464.

\bibitem{POURTAKDOUST2022134}
S.~H. Pourtakdoust, M.~F. Mehrjardi, M.~Hajkarim, Attitude estimation and
  control based on modified unscented kalman filter for gyro-less satellite
  with faulty sensors, Acta Astronautica 191 (2022) 134--147.

\bibitem{RN41}
J.~Hu, H.~Zhang, Bounded output feedback of rigid-body attitude via angular
  velocity observers, Journal of Guidance, Control, and Dynamics 36~(4) (2013)
  1240--1248.

\bibitem{G60189}
H.~C. Kjellberg, E.~G. Lightsey, Discretized constrained attitude pathfinding
  and control for satellites, Journal of Guidance, Control, and Dynamics 36~(5)
  (2013) 1301--1309.

\bibitem{2002Rapid}
B.~Wie, D.~Bailey, C.~Heiberg, Rapid multitarget acquisition and pointing
  control of agile spacecraft, Journal of Guidance Control Dynamics 25~(1)
  (2002) 96--104.

\bibitem{CHENG201861}
Y.~Cheng, D.~Ye, Z.~Sun, S.~Zhang, Spacecraft reorientation control in presence
  of attitude constraint considering input saturation and stochastic
  disturbance, Acta Astronautica 144 (2018) 61--68.

\bibitem{kulumani2017constrained}
S.~Kulumani, T.~Lee, Constrained geometric attitude control on so (3),
  International Journal of Control, Automation and Systems 15~(6) (2017)
  2796--2809.

\bibitem{HUA2024108738}
B.~Hua, J.~He, H.~Zhang, Y.~Wu, Z.~Chen, Spacecraft attitude reorientation
  control method based on potential function under complex constraints,
  Aerospace Science and Technology 144 (2024) 108738.

\bibitem{6978863}
U.~{Lee}, M.~{Mesbahi}, Feedback control for spacecraft reorientation under
  attitude constraints via convex potentials, IEEE Transactions on Aerospace
  and Electronic Systems 50~(4) (2014) 2578--2592.

\bibitem{G003606}
Q.~Hu, B.~Chi, M.~R. Akella, Anti-unwinding attitude control of spacecraft with
  forbidden pointing constraints, Journal of Guidance, Control, and Dynamics
  42~(4) (2019) 822--835.

\bibitem{SHEN2018157}
Q.~Shen, C.~Yue, C.~H. Goh, B.~Wu, D.~Wang, Rigid-body attitude stabilization
  with attitude and angular rate constraints, Automatica 90 (2018) 157 -- 163.

\bibitem{KALABIC2017293}
U.~V. Kalabić, R.~Gupta, S.~D. Cairano, A.~M. Bloch, I.~V. Kolmanovsky, {MPC}
  on manifolds with an application to the control of spacecraft attitude on
  {SO}(3), Automatica 76 (2017) 293 -- 300.

\bibitem{GJCD_DAE}
D.~Y. Lee, R.~Gupta, U.~V. Kalabić, S.~Di~Cairano, A.~M. Bloch, J.~W. Cutler,
  I.~V. Kolmanovsky, Geometric mechanics based nonlinear model predictive
  spacecraft attitude control with reaction wheels, Journal of Guidance,
  Control, and Dynamics 40~(2) (2017) 309--319.

\bibitem{8412335}
M.~M. {Nicotra}, E.~{Garone}, The explicit reference governor: A general
  framework for the closed-form control of constrained nonlinear systems, IEEE
  Control Systems Magazine 38~(4) (2018) 89--107.

\bibitem{7244340}
E.~{Garone}, M.~M. {Nicotra}, Explicit reference governor for constrained
  nonlinear systems, IEEE Transactions on Automatic Control 61~(5) (2016)
  1379--1384.

\bibitem{8890837}
M.~M. {Nicotra}, D.~{Liao-McPherson}, L.~{Burlion}, I.~V. {Kolmanovsky},
  Spacecraft attitude control with nonconvex constraints: An explicit reference
  governor approach, IEEE Transactions on Automatic Control 65~(8) (2020)
  3677--3684.

\bibitem{8675479}
M.~Hosseinzadeh, E.~Garone, An explicit reference governor for the intersection
  of concave constraints, IEEE Transactions on Automatic Control 65~(1) (2020)
  1--11.

\bibitem{Nicotra2016ConstrainedCO}
M.~Nicotra, E.~Garone, Constrained control of nonlinear systems: The explicit
  reference governor and its application to unmanned aerial vehicles, Ph.D.
  thesis, Universite Libre de Bruxelles (2016).

\bibitem{CHI2024108874}
B.~Chi, Q.~Hu, Saturated explicit reference governor for spacecraft constrained
  attitude reorientation control, Aerospace Science and Technology 145 (2024)
  108874.

\bibitem{HASAN2023}
M.~N. Hasan, Y.~Chen, J.~Liang, A.~Wen, Fixed-time fault-tolerant attitude
  control for flexible spacecraft without angular velocity sensor, ISA
  Transactions (2023, in press).

\bibitem{Yang2016Immersion}
S.~Yang, M.~R. Akella, F.~Mazenc, Immersion and invariance observers for
  gyro-free attitude control systems, Journal of Guidance Control Dynamics
  39~(11) (2016) 2567--2574.

\bibitem{ESPINDOLA2022377}
E.~Espíndola, Y.~Tang, A new angular velocity observer for attitude tracking
  of spacecraft, ISA Transactions 130 (2022) 377--388.

\bibitem{RN42}
R.~Schlanbusch, E.~I. Ingar~Grotli, Hybrid certainty equivalence control of
  rigid bodies with quaternion measurements, IEEE Transactions on Automatic
  Control 60~(9) (2015) 2512--2517.

\bibitem{6832520}
J.~G. {Romero}, R.~{Ortega}, I.~{Sarras}, A globally exponentially stable
  tracking controller for mechanical systems using position feedback, IEEE
  Transactions on Automatic Control 60~(3) (2015) 818--823.

\bibitem{G004302}
Q.~Dang, H.~Gui, H.~Wen, Dual-quaternion-based spacecraft pose tracking with a
  global exponential velocity observer, Journal of Guidance, Control, and
  Dynamics 42~(9) (2019) 2106--2115.

\bibitem{G002129}
Q.~Shen, C.~Yue, C.~H. Goh, Velocity-free attitude reorientation of a flexible
  spacecraft with attitude constraints, Journal of Guidance, Control, and
  Dynamics 40~(5) (2017) 1293--1299.

\bibitem{AIAAbook}
J.~L. Junkins, H.~Schaub, Analytical mechanics of space systems, American
  Institute of Aeronautics and Astronautics, 2009.

\bibitem{GUI20155832}
H.~Gui, G.~Vukovich, Adaptive integral sliding mode control for spacecraft
  attitude tracking with actuator uncertainty, Journal of the Franklin
  Institute 352~(12) (2015) 5832 -- 5852.

\bibitem{G002873}
M.~Diaz~Ramos, H.~Schaub, Kinematic steering law for conically constrained
  torque-limited spacecraft attitude control, Journal of Guidance, Control, and
  Dynamics 41~(9) (2018) 1990--2001.

\bibitem{661611}
A.~{Bemporad}, Reference governor for constrained nonlinear systems, IEEE
  Transactions on Automatic Control 43~(3) (1998) 415--419.

\end{thebibliography}





\end{document}